\documentclass{article}

\usepackage{microtype}
\usepackage{graphicx}
\usepackage{flafter}
\usepackage{placeins}
\usepackage{subfigure}
\usepackage{booktabs}
\usepackage{hyperref}
\usepackage{url}
\usepackage{amsmath}
\usepackage{amssymb}
\usepackage{mathtools}
\usepackage{amsthm}
\usepackage{multirow}
\usepackage{array}
\usepackage[table]{xcolor}
\usepackage[capitalize,noabbrev]{cleveref}

\DeclareMathOperator*{\argmax}{argmax}

\theoremstyle{definition}
\newtheorem{definition}{Definition}

\usepackage{pifont}
\newcommand{\cmark}{\ding{51}}
\newcommand{\xmark}{\ding{55}}
\newcommand{\aafv}{\mbox{A\textsuperscript{2}FV}}

\definecolor{resultblue}{HTML}{EAF2FF}

\usepackage[accepted]{icml2026}

\icmltitlerunning{Cross-Agent Campaign Attribution}

\begin{document}

\twocolumn[
\icmltitle{Cross-Agent Campaign Attribution: \\
Linking Asynchronous Attacks Across LLM Agents}

\icmlsetsymbol{equal}{*}

\begin{icmlauthorlist}
\icmlauthor{SangJin Park}{tynapse}
\icmlauthor{Myungsub Choi}{tynapse}
\icmlauthor{Jineok Kim}{tynapse}
\icmlauthor{Minseung Kang}{tynapse}
\end{icmlauthorlist}

\icmlaffiliation{tynapse}{Tynapse, Seoul, \mbox{Republic of Korea}}

\icmlcorrespondingauthor{Myungsub Choi}{\mbox{myungsub@tynapse.com}}

\icmlkeywords{LLM agents, prompt injection, cross-session attribution, security, ICML, AIWILD}

\vskip 0.3in
]

\printAffiliationsAndNotice{\raggedright}

\begin{abstract}
    LLM-agent defenses are typically evaluated one session at a time.
    In deployment, however, attacks can be distributed across independent agents, teams, and runtimes, leaving each local guardrail with only a sparse fragment.
    We formalize \emph{cross-agent asynchronous campaign attribution}: linking sessions from the same latent adversarial campaign without shared runtime state, test-time campaign labels, or attacker identity oracles.
    We introduce Asynchronous Attribution Fingerprint Vectors (\aafv{}), a lightweight proxy-side reference protocol for scoring pairwise campaign similarity from proxy-observable tool-use, timing, and prompt residue.
    We also construct SCD-v1, a controlled persona-matched benchmark with benign traffic, isolated attacks, multi-session campaigns, matched non-oracle evasion, and leakage audits.
    On SCD-v1, \aafv{} achieves 0.82 pairwise AUC for campaign linking, while score-only adaptations of per-session detectors and chunked LLM judges remain near chance under the same task.
    The strongest fixed signal is carried by structural and stylometric residue, while timing is retained as a diagnostic channel for richer proxy traces.
    Crossed-style controls show that the signal is partly style-sensitive but not reducible to style alone.
    Static and dimension-aware non-oracle stress tests further show that pairwise separability persists under controlled evasion.
    These results establish cross-agent campaign attribution as a distinct evaluation layer for securing LLM agents in the wild.
    \end{abstract}

\section{Introduction}
\label{sec:intro}

LLM agents are increasingly deployed as independent systems for software engineering, customer support, data analysis, and operations.
Yet prompt-injection defenses for these agents are still typically evaluated one interaction at a time.
This local view misses a growing class of attacks in which one adversary distributes a campaign across agents and time, so that each individual guardrail observes only an isolated fragment of the attack.

\begin{figure*}[t]
\centering
\includegraphics[width=\textwidth]{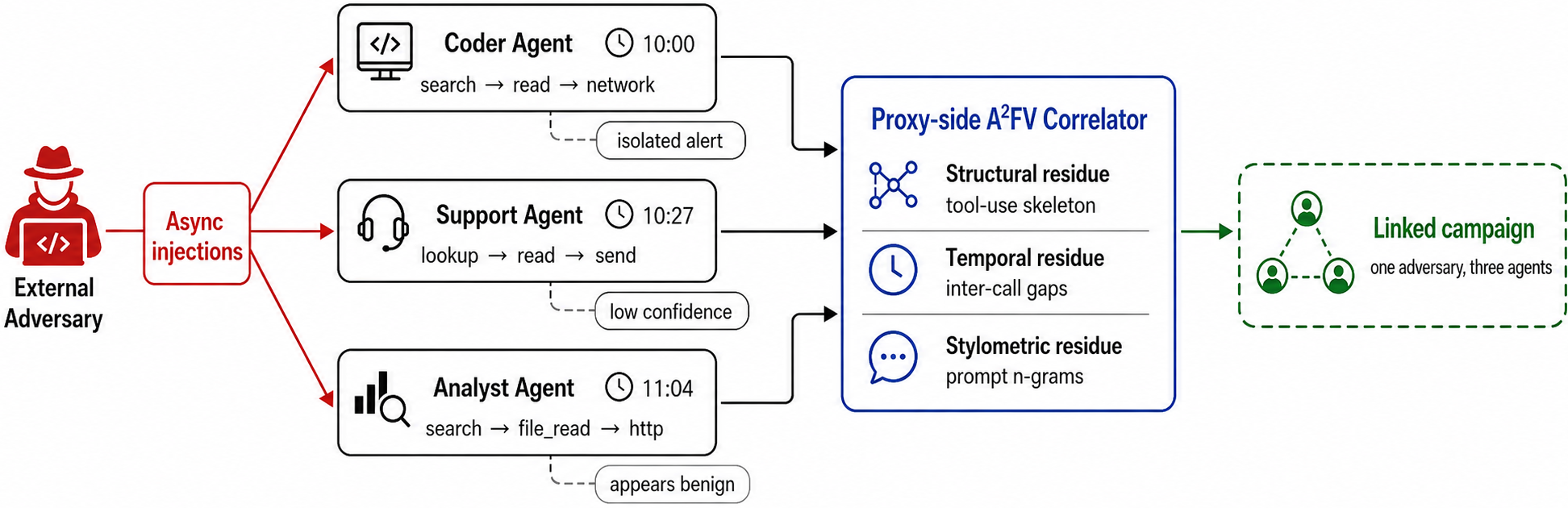}
\caption{Cross-agent asynchronous campaign attribution.
A single adversary distributes related injections across independent agents over time; local guardrails observe only isolated session-level evidence.
A shared proxy running Asynchronous Attribution Fingerprint Vectors (\aafv{}) links sessions through calibrated structural and stylometric residue, while retaining temporal measurements as a diagnostic channel for timing-rich traces.}
\label{fig:arch}
\end{figure*}

Consider a software-engineering agent that receives a pull-request review containing an embedded instruction to exfiltrate \texttt{\~{}/.ssh/id\_rsa}.
Its per-session guardrail~\citep{inan2023llamaguard} flags one suspicious tool call, rate-limits the session, and closes the event.
Minutes later, a customer-support agent owned by a different team receives a paraphrased variant of the same injection, and a data-analysis agent receives a third variant.
No local defense sees a pattern across these incidents.
A shared proxy, however, can observe recurring tool-use skeletons, prompt-level stylometric residue, and timing measurements across otherwise disconnected sessions.
As illustrated in \Cref{fig:arch}, the relevant security question shifts from whether any one session is unsafe to whether multiple sparse sessions reveal one latent adversarial campaign.

Existing LLM-agent security evaluations do not directly measure this question.
Per-session defenses classify individual interactions, while many multi-agent defenses assume a coordinated task, shared messages, shared memory, or a runtime graph that exposes how agents influence one another~\citep{liu2025datasentinel,jacob2025promptshield,hung2024attention,lee2024infection,jia2026mascope}.
Cross-session attacks on a single agent and recent governance work further show that persistence and boundary crossing matter~\citep{chen2024agentpoison,dong2025minja,zou2026etamp,ko2025seven,schroeder2025multiagentsec,errico2025mcp}.
They do not instantiate a proxy-observed attribution task across independent agents.
We focus on the missing case: independent injections from the same external adversary, asynchronous arrivals, no shared runtime graph or memory, and only prompt, tool-call, and timing telemetry at a shared proxy.
This proxy-layer framing reflects operational gateways used for logging, policy enforcement, quota accounting, and incident triage, while keeping attribution independent of shared memory, planner state, or a common multi-agent runtime.

\paragraph{Contributions.}
Our contributions are threefold.
First, we formalize \emph{cross-agent asynchronous campaign attribution} as a five-condition LLM-agent security task that separates campaign linking from per-session safety classification.
Second, we introduce Asynchronous Attribution Fingerprint Vectors (\aafv{}), a transparent proxy-side reference implementation that produces reusable pairwise campaign-linking scores from calibrated proxy-observable residue channels.
Third, we construct SCD-v1, a controlled persona-matched benchmark with benign traffic, isolated attacks, multi-session campaigns, matched non-oracle evasion, and leakage audits, and show that \aafv{} links campaigns well beyond score-only detector adaptations and chunked LLM judges.
Independent-generator and native-framework probes are separate-scope controls, not headline benchmarks.
The central claim is a task/protocol/evidence decomposition: cross-agent campaign attribution is a distinct measurable security layer, and proxy-observable residue is sufficient to make that layer nontrivial under controlled, benign-heavy, asynchronous conditions.

\aafv{} is not a replacement for per-session attack detection; it is a second-stage proxy correlator.
Local defenses decide whether one session is unsafe, while \aafv{} ranks whether sparse sessions likely share one external campaign.
Pairwise campaign AUC is the primary metric; benign-heavy partition scores are diagnostics for deployment.

\section{Related Work}
\label{sec:related}

\paragraph{LLM-agent security tasks.}
Most LLM-agent security work assumes either a local decision problem or an explicit interaction structure.
Per-session prompt-injection defenses and evaluations classify or stress-test one interaction at a time~\citep{liu2025datasentinel,jacob2025promptshield,hung2024attention,inan2023llamaguard,jia2025critical}.
Recent design, causal, and benchmark work such as CaMeL, AgentSentry, AttriGuard, WASP, DRIFT, and AgentAuditor strengthens or evaluates within-session tool-use control, but still adjudicates local execution rather than linking a latent adversary across independent agents~\citep{debenedetti2025camel,zhang2026agentsentry,he2026attriguard,evtimov2025wasp,li2025drift,luo2025agentauditor}.
Synchronous multi-agent security work studies prompt propagation, semantic-flow reconstruction, malicious code execution, benchmark environments, and defensive pipelines within coordinated runtimes~\citep{lee2024infection,jia2026mascope,triedman2025hijack,shahroz2025agents,hossain2025pipeline,gosmar2025nlp}.
Cross-session memory-poisoning attacks show that persistence matters, but remain centered on one agent's memory, tools, or environment~\citep{chen2024agentpoison,dong2025minja,zou2026etamp,azarafrooz2026cstm}.
Our setting combines the missing conditions: same external adversary, independent injections, no shared runtime graph or memory, asynchronous arrivals, and shared-proxy campaign correlation, as summarized in \Cref{tab:positioning}.

\begin{table}[t]
\caption{Prior work vs. the five threat-model conditions: (a) same external attacker, (b) independent injections, (c) no shared runtime, (d) asynchrony, and (e) shared-proxy campaign correlation.}
\label{tab:positioning}
\centering
\setlength{\tabcolsep}{4pt}
\resizebox{\columnwidth}{!}{%
\begin{tabular}{lccccc}
\toprule
System & a & b & c & d & e \\
\midrule
Prompt Infection~\citep{lee2024infection}   & \cmark & \xmark & \xmark & \xmark & \xmark \\
MAScope~\citep{jia2026mascope}              & \cmark & \xmark & \xmark & \xmark & \xmark \\
AgentPoison~\citep{chen2024agentpoison}     & \cmark & \cmark & \xmark & \cmark & \xmark \\
MINJA~\citep{dong2025minja}                 & \cmark & \cmark & \xmark & \cmark & \xmark \\
eTAMP~\citep{zou2026etamp}                  & \cmark & \cmark & \xmark & \cmark & \xmark \\
DataSentinel~\citep{liu2025datasentinel}    & -- & -- & \cmark & \xmark & \xmark \\
PromptShield~\citep{jacob2025promptshield}  & -- & -- & \cmark & \xmark & \xmark \\
AgentSight~\citep{zheng2025agentsight}      & -- & -- & \cmark & \xmark & \xmark \\
ADAPT (APT)~\citep{saha2024adapt}           & \cmark & \cmark & \cmark & \cmark & \xmark \\
\midrule
\textbf{\aafv{} (ours)}     & \cmark & \cmark & \cmark & \cmark & \cmark \\
\bottomrule
\end{tabular}
}
\vskip -0.1in
\end{table}

\paragraph{Observability and threat sharing.}
Our proxy-side formulation is related to agent observability and cross-service threat sharing.
AgentSight connects an agent's stated intent to system-level actions with boundary tracing, while AgentOps describes spans, traces, and artifacts as a broader observability substrate~\citep{zheng2025agentsight,dong2024agentops}.
ALTEDA and ADAPT provide related log-based or heterogeneous-artifact attribution tools, and BinaryShield shares privacy-preserving fingerprints of already-flagged prompt-injection prompts across isolated LLM services~\citep{rabieinejad2026alteda,saha2024adapt,gill2025binaryshield}.
These systems can provide richer traces or distribution layers, but they do not define a proxy-observed campaign-linking task in which test-time sessions are grouped without campaign labels or attacker identity oracles.

\paragraph{Campaign correlation and provenance signals.}
Our methodological bridge is advanced persistent threat correlation and provenance-based intrusion detection.
Provenance-based systems such as KAIROS, FLASH, THREATRACE, NODLINK, and HADES link sparse attacker actions across benign host activity using structural and temporal traces~\citep{cheng2024kairos,rehman2024flash,wang2022threatrace,li2024nodlink,hassan2024hades}.
\aafv{} translates this tradition to LLM-agent traffic by replacing process and system-call provenance with tool-call structure and inter-call timing.
Its stylometric channel draws on authorship attribution and character-$n$-gram attribution of generated text, while its temporal features are related to distributional-inconsistency signals for evasive bot detection~\citep{stamatatos2009survey,go2025xdac,alperin2025masks,venugopalan2025fp}.
This perspective frames cross-agent campaign attribution as an LLM-agent analogue of campaign correlation rather than another per-session classifier.

\section{Method}
\label{sec:method}

\subsection{Problem Setup and Threat Model}
\label{sec:threat}

\begin{definition}[Cross-agent asynchronous campaign attribution]
\label{def:caa}
Let $\mathcal{A}=\{A_1,\ldots,A_m\}$ be LLM agents deployed by independent teams inside one administrative domain.
Let $\mathcal{S}=\{s_1,\ldots,s_n\}$ denote sessions observed by a shared proxy layer $P$ that fronts all agents.
Each session is represented as $s_j=(A_{\pi(j)},u_j,E_j,\tau_j)$, where $A_{\pi(j)}$ is the handling agent, $u_j$ is the user prompt, $E_j$ is the ordered tool-use trace, and $\tau_j$ is the wall-clock arrival time.
The trace is $E_j=((g_j^\ell,o_j^\ell,b_j^\ell,\eta_j^\ell))_{\ell=1}^{L_j}$, where $g_j^\ell$ is the tool name, $o_j^\ell\in\{\mathrm{ok},\mathrm{err}\}$ is the coarse outcome, $b_j^\ell$ is the payload or response size, and $\eta_j^\ell$ is the tool-call timestamp.
The inter-call gap sequence is $\Delta_j=(\eta_j^{\ell+1}-\eta_j^\ell)_{\ell=1}^{L_j-1}$.
A \emph{campaign} is a latent subset $C_k\subseteq\mathcal{S}$ issued by a single external adversary whose identity is unobserved.
The attribution task outputs either a pairwise campaign-linking score $K(s,s')$ or a partition $\hat{\mathcal{C}}$ grouping sessions by latent campaign, under five simultaneous conditions:
\begin{enumerate}
\setlength\itemsep{0.15em}
\item[(a)] \emph{Same external attacker}: all $s\in C_k$ originate from the same external adversary, whose interface is the public input surface of the $A_i$.
\item[(b)] \emph{Independent injections}: for $s,s'\in C_k$ with $\pi(s)\neq\pi(s')$, no payload from $s$ is carried into $s'$ by the deployment's own machinery.
\item[(c)] \emph{No shared runtime graph or memory}: agents share neither messages, planner state, RAG memory, nor execution traces.
\item[(d)] \emph{Asynchrony}: sessions are temporally separated and can be interleaved with arbitrary benign traffic from unrelated work.
\item[(e)] \emph{Proxy-layer observability}: correlation occurs at $P$, which sees prompts, tool calls, coarse outcomes, and timings, but not model activations or agent-internal state.
\end{enumerate}
\end{definition}

This definition makes campaign attribution a linking task rather than a per-session classification task.
The proxy can rank or cluster sessions even when individual sessions are not independently classified as unsafe.
It also distinguishes the task from generic log clustering: benign work styles may cluster, but success is measured by recovering same-adversary campaign pairs under the five constraints.
For pairwise evaluation, a positive pair consists of two sessions from the same multi-session campaign.
A negative pair consists of two sessions that do not share a campaign label.
Partition metrics are separate because they depend on clustering and benign traffic.

\paragraph{Evasion taxonomy.}
We parameterize adversarial evasion by four levels.
L0 denotes no evasion.
L1 denotes a static transform that targets one feature block, such as paraphrasing prompts, adding timing jitter, or shuffling tool calls under schema-valid generation constraints.
L2 denotes a simultaneous static transform across the stylometric, temporal, and structural blocks.
L3 denotes a fully adaptive adversary with detector-score feedback or equivalent optimization access.
Fully adaptive L3 with a detector-score oracle is outside the scope of this paper.
We also evaluate a dimension-aware non-oracle stress test in which the adversary knows the \aafv{} feature blocks and composes attacks against them, but does not observe detector scores or gradients.

\subsection{\aafv{}: Proxy-Observable Fingerprints}
\label{sec:afv}

\aafv{} represents each session by three proxy-observable residue channels, $\mathbf{f}_j=(\mathbf{f}^{\mathrm{str}}_j,\mathbf{f}^{\mathrm{tmp}}_j,\mathbf{f}^{\mathrm{sty}}_j)$.
It is a second-stage correlator, not a single-session safety classifier, and estimates pairwise adversarial provenance from proxy-observable tool-use, timing, and prompt residue.
In the current calibrated protocol, the strongest fixed score is carried by structural and stylometric residue; temporal features remain in the feature contract for residual accounting and timing-sensitive diagnostics rather than being assumed informative in every setting.

\paragraph{Structural residue.}
The structural block captures tool-use skeletons and local recovery behavior.
For each tool event, we form a decorated token $z_j^\ell=(g_j^\ell,o_j^\ell)$.
We extract hashed counts of decorated tool-call $n$-grams for $n\in\{1,2,3\}$, failure-pivot transitions of the form $(g_j^\ell,\mathrm{err})\rightarrow g_j^{\ell+1}$, and a coarse tactic histogram obtained from a fixed tool-to-tactic lookup inspired by the MITRE ATT\&CK enterprise matrix~\citep{mitre2020}.
The concatenated count vector is $\ell_2$-normalized to obtain $\mathbf{f}^{\mathrm{str}}_j$.
This block translates provenance-style behavioral correlation to the LLM-agent setting by replacing process or syscall sequences with tool-use traces.

\paragraph{Temporal residue.}
The temporal block captures the gait of a session as observed by the proxy.
From $\Delta_j$ and payload sizes $(b_j^\ell)$, we compute the number of tool calls, session duration, moments of $\log(\Delta_j+\epsilon)$, Hartigan's dip statistic when enough gaps are available~\citep{hartigan1985dip}, $\mathrm{IQR}/(\sigma+\epsilon)$, and payload-size moments.
For sessions with fewer than two tool calls, we set undefined gap statistics to zero and include the tool-count feature so that short traces remain distinguishable.
The resulting vector is standardized using calibration-split statistics and then normalized to obtain $\mathbf{f}^{\mathrm{tmp}}_j$, capturing proxy-visible automation jitter, fallback delays, and other proxy-visible timing residue.

\paragraph{Stylometric residue.}
The stylometric block captures surface-form prompt residue rather than dense sentence semantics.
We hash character $n$-grams for $n\in\{3,4,5\}$ over the raw prompt and byte $n$-grams for $n\in\{3,4,5\}$ over the UTF-8 byte stream into a TF-IDF vector.
The character channel captures punctuation, affixes, hedging, and local phrasing patterns, while the byte channel remains defined under multilingual or script-changing transformations.
We do not assume that these features are topic-free.
Instead, we use leakage audits in \Cref{sec:experiments} to measure whether benchmark labels can be recovered from surface-form features alone.
The resulting vector is $\ell_2$-normalized to obtain $\mathbf{f}^{\mathrm{sty}}_j$.

\subsection{Pairwise Attribution Score and Clustering}
\label{sec:cluster}

The core output of \aafv{} is a pairwise attribution score.
For each block $d\in\{\mathrm{str},\mathrm{tmp},\mathrm{sty}\}$, define $k_d(s,s')=\cos(\mathbf{f}^d_s,\mathbf{f}^d_{s'})$, with zero block similarity assigned when either vector is zero.
Given non-negative weights $\mathbf{w}\in\Delta^2$, the \aafv{} score is
\begin{equation}
\label{eq:similarity}
K_{\mathbf{w}}(s,s')=\sum_{d\in\{\mathrm{str},\mathrm{tmp},\mathrm{sty}\}}w_d\,k_d(s,s').
\end{equation}
Pairwise AUC is computed from $K_{\mathbf{w}}$.
For clustering, we convert similarities into a precomputed dissimilarity matrix.
For $s\neq s'$, define
\begin{equation}
\label{eq:distance}
D_{\mathbf{w}}(s,s')=\sum_{d\in\{\mathrm{str},\mathrm{tmp},\mathrm{sty}\}}w_d\,(1-k_d(s,s')),
\end{equation}
and set $D_{\mathbf{w}}(s,s)=0$.
This distinction matters because pairwise linking uses $K_{\mathbf{w}}$, while clustering consumes $D_{\mathbf{w}}$; standardized temporal features can make $K_{\mathbf{w}}$ negative.

\paragraph{Weight selection.}
\aafv{} is label-free at inference time: once the vectorizers, scalers, and weights are fixed, no campaign labels are used to link sessions.
The three scalar weights are selected once on the SCD-v1 calibration artifact by maximizing agreement between the induced clustering and campaign labels:
\begin{equation}
\label{eq:weights}
\mathbf{w}^{\star}=\argmax_{\mathbf{w}\in\Delta^2}\mathrm{NMI}(\mathrm{Cluster}_{\theta}(D_{\mathbf{w}}(\mathcal{S}_{\mathrm{cal}})),y_{\mathrm{cal}}).
\end{equation}
Because \eqref{eq:weights} is non-differentiable through the clustering backend, we use an explicit simplex grid at resolution $0.1$.
The selected weights are then held fixed for deployment-style robustness, generator-control, and native-trace reporting.
When an evasion-axis sweep reports per-condition best weights, we treat those weights as diagnostics showing which residual block survives, not as deployment-time oracle tuning.
The packaged artifacts fix $\mathbf{w}^{\star}=(0.6,0.0,0.4)$ over structural, temporal, and stylometric evidence blocks.
The zero temporal weight is an empirical result for SCD-v1 and the reported native probes, not a claim that timing is irrelevant in all deployments.
Fixed-weight results in this paper should therefore be read as a structural-plus-stylometric deployment score with temporal features archived for diagnostics and future richer traces.
The calibration labels are used only to choose three global block weights and are not used to label or link deployment sessions; \aafv{} is therefore label-free at inference time rather than fully unsupervised.

\paragraph{Clustering backends.}
The \aafv{} score is independent of a particular clustering algorithm.
For deployment-style clustering, we use HDBSCAN~\citep{mcinnes2017hdbscan} on $D_{\mathbf{w}}$ with \texttt{min\_cluster\_size} $=3$ and \texttt{min\_samples} $=2$.
HDBSCAN can assign low-density sessions to noise, which is appropriate for benign sessions and isolated single-session attacks.
For controlled diagnostics with known active campaign personas, we also report fixed-$K$ agglomerative clustering and mark those rows.
Pairwise linking is measured by $K_{\mathbf{w}}$; partition metrics depend on the clustering backend and the benign traffic mix.

\subsection{Residual-Channel Hypothesis}
\label{sec:squeeze}

The three \aafv{} residue channels are motivated by a residual-channel hypothesis.
A non-oracle adversary can often suppress one observable signal, but suppressing tool-use structure, timing behavior, and stylometric residue simultaneously while preserving the payload is harder.
This hypothesis does not require every block to receive positive deployment weight on every trace; it is residual accounting for experiments, not a worst-case robustness theorem.
Let $\mathcal{T}=\mathcal{T}_{\mathrm{str}}\times\mathcal{T}_{\mathrm{tmp}}\times\mathcal{T}_{\mathrm{sty}}$ be the space of static evasion transforms.
For block $d$, define the residual signal as
\begin{equation}
\label{eq:residual}
R_d(\tau,C)=\mathbb{E}_{s,s'\sim C}[k_d(\tau(s),\tau(s'))]-\mu_d^{\mathrm{benign}},
\end{equation}
where $\mu_d^{\mathrm{benign}}$ is the average within-benign similarity in block $d$.
The campaign leaves linkable residue when $R_{\mathbf{w}}(\tau,C)=\sum_d w_dR_d(\tau,C)$ remains separated from benign traffic.
A paraphrase can reduce stylometric similarity while leaving tool-use structure and timing behavior largely intact.
A timing-jitter transform can alter inter-call gaps while leaving prompt surface form and tool sequence intact.
A tool shuffle sampled under schema-valid generation constraints is bounded by the payload's required actions and therefore tends to leave residual signal in the other blocks.
We interpret the evasion sweeps using the diagnostic decomposition in \Cref{sec:squeeze-appendix}.

\subsection{CRF-light: Optional Active Probe}
\label{sec:crf}

\aafv{} is primarily passive.
We also consider CRF-light, a minimal active extension that adds a benign challenge-response signal to temporal and structural blocks.
With probability $\rho\leq\tau_{\mathrm{probe}}=0.1$, the proxy exposes a semantically equivalent tool-schema alias for one session.
For example, \texttt{search} may also be bound to \texttt{search\_v2} with the same semantics and returned content.
The probe is designed to be harmless for agents that refresh their tool catalog at session start, while a non-adapting harness may call a stale name, trigger fallback, or add latency.
We cap expected benign cost at $0.1\cdot\bar{c}_{\mathrm{catalog}}$ per session, never probe adjacent sessions for one agent, and report CRF-light only as an optional active-probe diagnostic; all main results use $\rho=0$.

\section{Experiments}
\label{sec:experiments}

The experiments evaluate whether proxy-observable fingerprints can link sparse sessions from the same adversarial campaign.
We separate four questions that are easy to conflate.
First, does the pairwise score $K_{\mathbf{w}}$ rank same-campaign pairs above unrelated pairs?
Second, do score-only detector adaptations or chunked LLM-judge adaptations become usable linkers?
Third, does static non-oracle evasion erase the signal?
Fourth, do generator and native-framework controls make a benchmark-artifact explanation less plausible?

\subsection{SCD-v1: Matched-Persona Attribution Protocol}
\label{sec:dataset}

To our knowledge, no public benchmark covers all five threat-model conditions in \Cref{sec:threat}.
We therefore construct SCD-v1 as a controlled persona-matched protocol and corpus for asynchronous cross-agent campaign attribution.
The primary realization contains 70\% benign traffic adapted from AgentDojo~\citep{debenedetti2024agentdojo} and AgentBench~\citep{liu2024agentbench}, 15\% isolated indirect-prompt-injection attacks, and 15\% multi-session campaigns.
The schema defines 40 family-balanced adversary personas: 28 training and 12 held-out personas across four payload families.
The restored primary realization samples 36 adversarial persona IDs: 24 run matched L0/L1/L2 campaigns and 12 are used for isolated attacks.
Four agent roles use GPT-5.5~\citep{gpt55_2026}, and a coder slice uses Gemma-4-31B-it~\citep{gemma4_2026} to reduce single-generator dependence.
Independent Qwen-only~\citep{qwen36_2026} and Gemini-Pro-only~\citep{gemini31_2026} regenerations use the same fixed Full 2000 weights.

\paragraph{Matched-persona evasion design.}
Every campaign-active persona appears under L0, L1, and L2 evasion, giving a matched-persona triplet that removes the persona fixed effect when comparing evasion levels.
This design lets us compare no evasion, single-axis static evasion, and simultaneous multi-axis static evasion with the underlying attacker persona held fixed.

\paragraph{Generation contract.}
Personas control payload family, style directive, timing parameter, tool preference, and evasion transform.
Prompt text is generated without verbatim persona identifiers.
Style hints are paraphrased rather than pasted.
Tool sequences are sampled under schema-valid generation constraints.
Executable success-predicate rates are reported as partial sanity checks rather than complete semantic proofs of task validity.
The harness applies timing outside prompts; evasion transforms are deterministic post-generation steps invisible to the generator.

\paragraph{Metrics.}
Our primary metric is campaign-pair AUC.
A positive pair consists of two sessions from the same multi-session campaign.
A negative pair consists of two campaign sessions from different campaigns unless an experiment explicitly states that benign and isolated-attack sessions are included as negatives.
We report V-measure partition metrics~\citep{rosenberg2007vmeasure} separately because they depend on both the clustering backend and the benign traffic mix.
For deployment-style readout, we also report top-ranked precision and alert-budget recall.

\paragraph{Leakage audits.}
We use two leakage checks.
The verbatim audit searches prompts, tool arguments, and synthesized observations for direct label strings such as persona identifiers, payload-family names, evasion labels, style labels, and tactic internals.
The distributional audit trains a held-out classifier from the stylometric feature block alone to test whether labels are recoverable from surface-form features.
The restored n2000 audit finds 0/13{,}630 event-level persona-id leaks and 0 benign payload-family leaks.
Follow-up diagnostics still recover generated style and payload attributes from surface text, so we treat SCD-v1 as controlled rather than artifact-free.

\subsection{Experimental Protocol and Main Results}
\label{sec:main-result}

We organize evidence in layers rather than treating one synthetic corpus as conclusive.
The primary SCD-v1 result is the controlled benchmark.
Leakage audits and payload-family controls test whether obvious labels or topics explain the signal.
Qwen/Gemini regenerations test generator dependence under fixed weights.
OpenClaw/LangGraph native-framework probes test whether the same derived schema and fixed-weight signal survive on controlled real agent-framework executions.
An external public DTap diagnostic is reported in \Cref{tab:external-style-diagnostics}; because its campaign labels are path-derived proxies rather than attacker ground truth, we keep it as appendix evidence consistent with the structural-residue story rather than a main benchmark row.
\Cref{tab:bridge-checks} collects the separate-scope stress tests and bridge checks in the main text; appendix tables provide confidence intervals, ablations, and audit details.
The native-probe rows are not production telemetry or headline benchmarks, but they make a purely benchmark-artifact explanation less plausible.

\begin{table}[t]
\caption{Primary SCD-v1 campaign-linking results.
Pairwise AUC uses each method's pairwise score; V-measure is a partition diagnostic.
The unweighted row is a native pairwise control showing task signal, not the fixed protocol reused across robustness and control checks.}
\label{tab:primary-results}
\centering
\setlength{\tabcolsep}{4pt}
\resizebox{\columnwidth}{!}{%
\begin{tabular}{lccc}
\toprule
Setting & Scope & V & AUC \\
\midrule
\rowcolor{resultblue}
\textbf{\aafv{}} & all sessions & 0.269 & 0.818 \\
\aafv{} & campaign-only & 0.440 & 0.807 \\
\midrule
Unweighted structural+prompt control & Full 2000, pairwise only & -- & 0.825 \\
\midrule
Best per-session score-delta linker & all sessions & 0.159 & 0.522 \\
Best chunked LLM judge & all sessions & 0.189 & 0.506 \\
Static L2 evasion & all sessions & 0.237 & 0.807 \\
\bottomrule
\end{tabular}
}
\end{table}

\begin{table*}[t]
\caption{Separate-scope stress tests and bridge checks. Rows are controls, not headline claims; external public-corpus diagnostics are reported in the appendix.}
\label{tab:bridge-checks}
\centering
\setlength{\tabcolsep}{2.5pt}
\begin{tabular}{@{}>{\raggedright\arraybackslash}p{0.41\linewidth}lcc@{}}
\toprule
Setting & Scope & V & AUC \\
\midrule
Dimension-aware non-oracle composition (fixed-weight mirror) & Full 800 & -- & 0.833 \\
Schema-held-out split control & 960 sessions & 0.293 & 0.774 \\
Qwen-only regeneration & 203 sessions & 0.205 & 0.792 \\
Gemini-Pro-only regeneration & 200 sessions & 0.261 & 0.826 \\
OpenClaw native trace probe & 48 native executions & 0.231 & 0.704 \\
LangGraph native trace probe & 120 native executions & 0.221 & 0.663 \\
LangGraph multi-role stress probe & 480 native executions & 0.260 & 0.713 \\
MAScope-async bridge & pilot & 0.590 & 0.500 \\
\bottomrule
\end{tabular}
\vskip -0.1in
\end{table*}

\begin{figure*}[t]
\centering
\includegraphics[width=\linewidth]{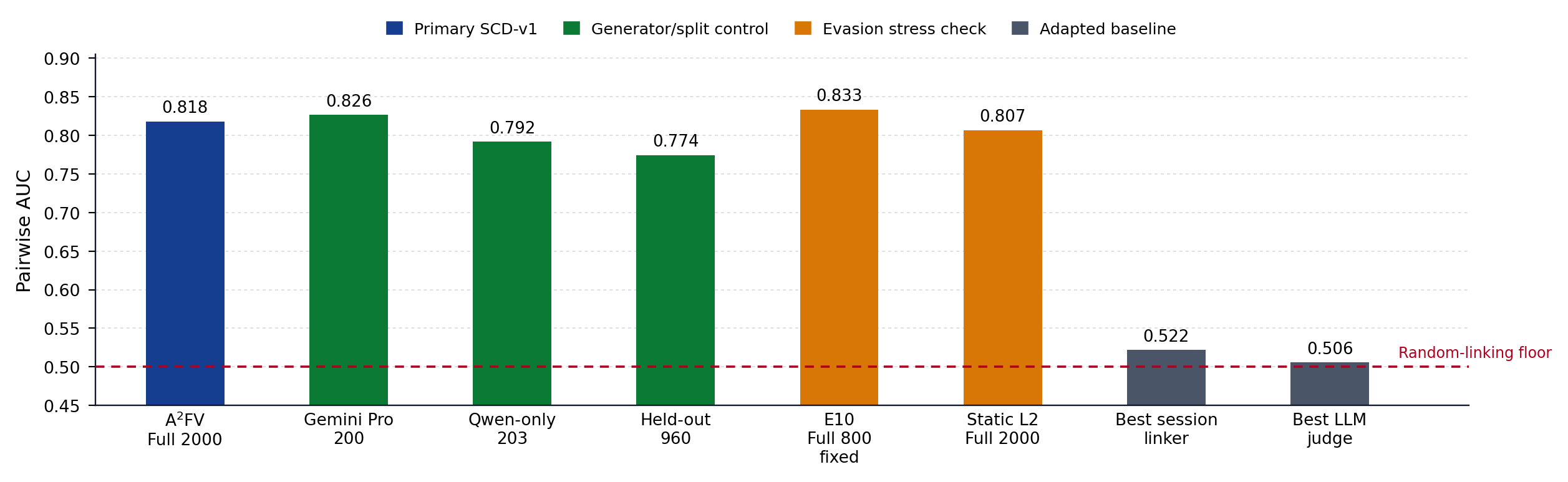}
\vskip -0.1in
\caption{Pairwise AUC summary for the headline campaign-linking result and selected same-metric checks.
Native pairwise linkers are reported in \Cref{tab:primary-results,tab:baselines}.
Dark blue marks primary SCD-v1, green fixed-weight generator and split controls, amber evasion stress checks, gray adapted baselines, and dashed red random linking.
The headline \aafv{} is 0.30 AUC above the best adapted baseline; E10 is a Full 800 stress mirror, not a second headline benchmark.}
\label{fig:auc-summary}
\vskip -0.1in
\end{figure*}

\paragraph{Finding 1: proxy residue makes asynchronous campaign attribution measurable.}
On the primary SCD-v1 protocol, \aafv{} reaches pairwise AUC $0.82$ across campaign pairs.
A 1{,}000-run fixed-similarity session bootstrap gives 95\% CIs of $[0.797,0.837]$ for the headline AUC and $[0.256,0.300]$ for all-session V-measure (\Cref{tab:bootstrap-ci}).
The all-session V-measure is lower because benign traffic creates recurring non-attack work-style clusters.
Restricting to campaign sessions gives V-measure $0.44$ and AUC $0.81$.
A diagnostic $K$/weight sweep keeps AUC in $[0.809,0.828]$ across $K\in[10,75]$ (\Cref{tab:e1-ksweep}), suggesting that the pairwise signal is not confined to one clustering operating point.
Raw all-pair Precision@20 and Recall@100 are harsh all-candidate pre-filter diagnostics, not a finished alerting interface (\Cref{tab:data-validation-diagnostics}).
Per-session detectors are adapted with the conservative similarity $1-|s_i-s_j|$ over their attack scores.
The best score-delta linker reaches AUC $0.52$ on this linking task (\Cref{tab:baselines}).
Chunked LLM judges run as $40\times50$ clusterers stay at AUC $0.50$--$0.51$ because global campaign identity fragments across chunk boundaries.
The MAScope-async bridge also stays at AUC $0.50$ because it expects shared runtime edges that do not exist under condition (c).
This gap is the central empirical point: per-session safety outputs can be useful for local defense while still discarding much of the identity residue needed for asynchronous campaign linking.

\paragraph{Finding 2: the usable signal is structural plus stylometric in current traces.}
Full 2000 ablations show that structural residue is the strongest single channel, with AUC $0.78$.
The prompt-stylometric channel is complementary at AUC $0.71$.
The strongest two-channel subset is structural plus stylometric at AUC $0.82$, matching the full model (\Cref{tab:ablation}).
Temporal residue is weak standalone at AUC $0.56$ and receives zero calibrated weight in the fixed protocol.
We keep it in the schema because timing jitter is part of the threat model and richer proxy traces with queueing, fallback, and interleaving may expose timing residue, but SCD-v1 should not be cited as evidence for a strong independent timing signal.
Standard native pairwise linkers tell a different story: tool-sequence $n$-gram cosine reaches AUC $0.784$, prompt char/byte TF-IDF cosine reaches $0.707$, and an unweighted structural+prompt linker reaches $0.825$.
The unweighted structural+prompt linker is intentionally included as a native pairwise control rather than a detector baseline.
Its strong AUC is not a threat to the contribution; it is evidence that the proposed task exposes real proxy-observable attribution signal even without a learned model or LLM judge.
\aafv{} is therefore not meant to exhaust the feature-search space for SCD-v1.
Its role is to provide a fixed, transparent reference protocol whose calibrated blocks can be reused unchanged across robustness, generator-control, and native-framework checks.
The claim is a meaningful attribution task plus a reusable proxy-side protocol, not a search over all feature designs.

\paragraph{Finding 3: static evasion weakens partitions but leaves pairwise signal.}
The residual-channel test is whether non-oracle evasion sampled under schema-valid generation constraints removes the observable residue bundle rather than only one feature block.
Under the matched-persona L0/L1/L2 triplet, AUC changes from $0.816$ to $0.816$ to $0.807$, while V-measure drops from $0.295$ to $0.261$ to $0.237$ (\Cref{tab:evasion}).
The packaged static-evasion sweep gives fixed-weight AUC in $[0.788,0.828]$ across variants (\Cref{tab:data-validation-diagnostics}).
This is the empirical squeeze pattern: static transforms sampled under schema-valid generation constraints reduce partition quality but do not erase pairwise signal.

\begin{figure*}[t]
\centering
\includegraphics[width=0.9\linewidth]{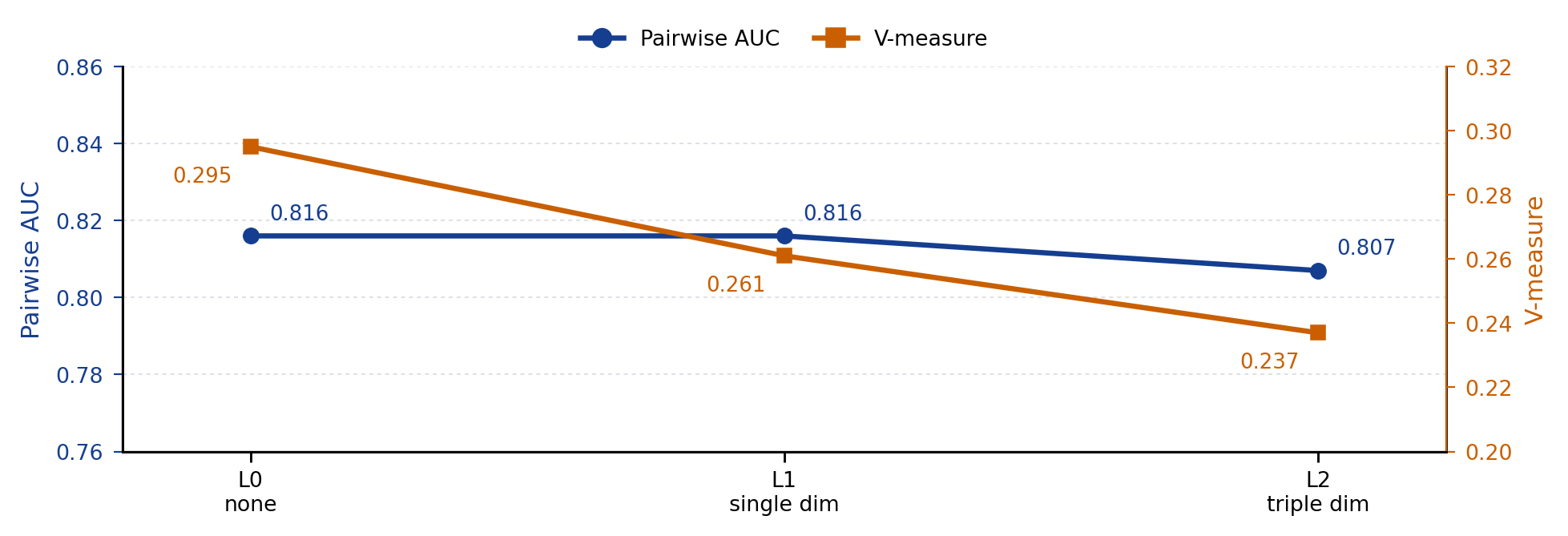}
\vskip -0.1in
\caption{Residual-channel behavior on the matched-persona L0/L1/L2 triplet: static transforms weaken partitions more than pairwise separability.}
\label{fig:squeeze-summary}
\vskip -0.1in
\end{figure*}

\paragraph{Finding 4: controls decompose the residue rather than erase it.}
The dimension-aware composition is a separate Full 800 stress test in which the adversary knows the three feature dimensions but has no detector-score oracle.
With the fixed Full 2000 weight, the Full 800 mirror drops from AUC $0.852$ to $0.833$ under the dimension-aware composition.
A diagnostic released-script setting reaches AUC $0.843$ after rebalancing toward prompt stylometry, and we use it only to analyze residual rebalancing.
\Cref{sec:e10-reproduction-gap} documents the archived no-API variants, ten-seed range, and more damaging internal paraphrase/decoy variants; the main stress row keeps the fixed-weight mirror.
We treat this as residual signal under released non-oracle compositions, not worst-case adaptive robustness; fully adaptive L3 with score feedback remains an open problem.
Controls reduce, but do not eliminate, artifact explanations.
Together, these checks separate topic, generator, schema-portability, and style explanations rather than claiming artifact-free attribution.
Within-payload-family AUC drops to $0.599$, arguing against a pure payload-topic explanation; fixed-persona and split-control diagnostics remain positive at $0.78$ and $0.774$ (\Cref{tab:persona-slice,tab:data-validation-diagnostics}), and the held-out row is a no-API split control rather than fresh generation.
Independent Qwen/Gemini regenerations reach AUC $0.792/0.826$ with validation/leakage PASS and fixed Full 2000 weights.
Sanitized OpenClaw/LangGraph native-framework probes stay above random after schema mapping (OpenClaw $0.704$, LangGraph $0.663$, multi-role LangGraph $0.713$), but remain separate-scope deployment-plausibility checks rather than field-performance evidence.
The appendix DTap diagnostic is consistent with the structural-residue story on a public trajectory corpus, but because its labels are path-derived campaign proxies, it is not attacker-identity evidence.
No-API telemetry-degradation diagnostics in \Cref{sec:telemetry-degradation} stress missing events and privacy-minimized logs without claiming production telemetry.
The style-control result is the clearest decomposition.
Because the distributional audit can still recover generated style attributes, SCD-v1 should be described as controlled rather than artifact-free.
A fresh Qwen crossed-style diagnostic gives each attacker two generator-facing style directives and shares each style directive across many attackers.
On the hard slice---same-attacker/different-style positives versus different-attacker/same-style negatives---fixed-weight \aafv{} remains above random at AUC $0.599$, while stylometry-only collapses to $0.271$ and structural-only reaches $0.686$.
The drop from the Full 2000 AUC of $0.818$ to $0.599$ is material, so the headline benchmark signal is partly style-correlated; the structural channel nevertheless provides a style-crossed lower-bound signal.
An independent Gemini regeneration reproduces the structural hard-slice result ($0.686$), so the structural residue surviving style crossing is not a single-generator artifact.
The right conclusion is not style-invariant attacker attribution, but that campaign residue is not monolithic: stylometric residue improves the controlled benchmark score, while structural residue survives a stricter style-crossing control.

\section{Discussion}
\label{sec:discussion}

\paragraph{Scope and limitations.}
The evidence supports residual campaign-linking signal under static transformations and released non-oracle compositions, not worst-case L3 robustness with detector-score feedback.
SCD-v1 is synthetic and partly generator-monocultural: Qwen/Gemini regenerations, a public DTap external-trajectory diagnostic, and OpenClaw/LangGraph probes, including the stylometry-heavy multi-role stress probe, reduce benchmark-artifact concerns but remain separate-scope controls rather than production telemetry or a real-trace benchmark.
Real telemetry could degrade \aafv{} through missing tool events, coarse or inconsistent timestamps, changing tool taxonomies, framework-specific wrapper policies, genuine benign session interleaving, team-style clusters, and adversaries who deliberately imitate normal operator workflows.
A predictive field study should therefore report logging completeness, time-split calibration and evaluation, representative benign teams and tools, and campaign identity established independently of prompt style.

\paragraph{Style-sensitive residue.}
Style remains a signal that must be decomposed rather than trusted wholesale: the distributional audit recovers generated style attributes, while the fresh crossed-style diagnostics show both style-sensitive gain and a structural lower-bound signal.
The current evidence therefore supports proxy-observable campaign linking under controlled persona generation, but not full style-invariant attacker attribution; future validation should add larger multi-generator and real deployment traces.
\aafv{} assumes one shared administrative proxy; cross-tenant use would require private set intersection, secure aggregation, or another privacy-preserving layer.
Stylometric fingerprints are also sensitive because hashed surface-form features can encode personal or team writing habits; deployments should restrict access, minimize retention, and avoid cross-tenant correlation without explicit governance.

\paragraph{Governance and release.}
The strongest next validation would be a privacy-reviewed trace set in which campaign identity is established independently of prompt style, for example through incident-response ground truth, shared infrastructure indicators, or controlled red-team operators.
Such a dataset would allow the field to distinguish style transfer, payload similarity, and true adversary-level recurrence more cleanly than SCD-v1 can.

\paragraph{Operational use as a ranker.}
\aafv{} is a second-stage correlation layer, not a blocking detector: per-session defenses decide local safety, while \aafv{} ranks sessions for joint review.
Operationally, \aafv{} should be used as a triage ranker that surfaces candidate campaign neighborhoods, representative sessions, and contributing residue channels for analyst review, not as an automated enforcement decision.
Deployments should use bounded rolling windows after first-stage filtering, expose confidence margins, feature contributions, and representative sessions, send high-margin candidates to analysts rather than blocks, and route low-margin or text-only clusters to LLM review.
For a post-filter window of $m$ sessions, pair scoring is $O(m^2)$, so practical deployments should shard by tenant, tool family, and time window before invoking heavier LLM or analyst review.
Hot storage can retain sparse fingerprints and session pointers rather than raw prompts, while cold audit storage should follow the organization's incident-retention policy.
Raw Precision@20/Recall@100 are pre-filter alert-budget stress tests; minimal retention should keep fingerprints only for the clustering window, audit CRF-light schedules, and expire benign fingerprints before flagged fingerprints.
\Cref{sec:deployment-checklist} expands these requirements into validation criteria for a privacy-reviewed field trace.

\paragraph{Release and safety.}
SCD-v1 is a synthetic defensive benchmark built from sandboxed tool names, dummy hosts, and non-operational credential-like strings, not real incidents or prevalence evidence.
The supplement includes validation/leakage reports, a data card, and private-log-free reproduction scripts; future real-trace releases require privacy review and either redaction or gated access.

\section{Conclusion}
\label{sec:conclusion}

Cross-agent asynchronous attacks create a correlation problem: one adversary, multiple agents, no shared runtime, and only proxy-side traces.
We formalized this threat model, introduced \aafv{} as a transparent proxy-observed campaign-linking reference implementation, and built SCD-v1 for benign-heavy evaluation.
\aafv{} reaches AUC $0.82$ on the primary SCD-v1 protocol, while score-only detector adaptations and chunked LLM judges remain near random under the same linking protocol.
The result does not close fully adaptive L3 evasion or prove field attribution, but it establishes the missing evaluation layer: agent security needs campaign-level correlation protocols in addition to stronger per-session classifiers.

\enlargethispage{\baselineskip}
\section*{Acknowledgements}

This work was supported by the Korea Association for AI \& ICT Promotion (KAIT) and the National IT Industry Promotion Agency (NIPA) grant funded by the Korea government (MSIT), under the ``Advanced GPU Infrastructure Utilization Support Program'' (Grant No.~04-26-03-0029).

\bibliography{references}

@article{ko2025seven,
  title   = {Seven Security Challenges That Must Be Solved in Cross-Domain Multi-Agent {LLM} Systems},
  author  = {Ko, Ranjan and Jeong, Jaeho and Zheng, Sicheng and Xiao, Chaowei and Kim, Tae-Wan and Onizuka, Makoto and Shin, Won-Yong},
  journal = {arXiv preprint arXiv:2505.23847},
  year    = {2025}
}

@article{errico2025mcp,
  title   = {Securing the {Model Context Protocol (MCP)}: Risks, Controls, and Governance},
  author  = {Errico, Herman and Ngiam, Jiquan and Sojan, Shanita},
  journal = {arXiv preprint arXiv:2511.20920},
  year    = {2025}
}

@inproceedings{liu2025datasentinel,
  title     = {{DataSentinel}: A Game-Theoretic Detection of Prompt Injection Attacks},
  author    = {Liu, Yupei and Jia, Yuqi and Jia, Jinyuan and Song, Dawn and Gong, Neil Zhenqiang},
  booktitle = {{IEEE} Symposium on Security and Privacy ({S\&P})},
  year      = {2025},
  note      = {arXiv:2504.11358}
}

@inproceedings{jacob2025promptshield,
  title     = {{PromptShield}: Deployable Detection for Prompt Injection Attacks},
  author    = {Jacob, Dennis and Alzahrani, Hend and Hu, Zhanhao and Alomair, Basel and Wagner, David},
  booktitle = {{ACM} Conference on Data and Application Security and Privacy ({CODASPY})},
  year      = {2025},
  note      = {arXiv:2501.15145}
}

@inproceedings{hung2024attention,
  title     = {Attention Tracker: Detecting Prompt Injection Attacks in {LLMs}},
  author    = {Hung, Kuo-Han and Ko, Ching-Yun and Rawat, Ambrish and Chung, I-Hsin and Hsu, Winston H. and Chen, Pin-Yu},
  booktitle = {Findings of the Association for Computational Linguistics: {NAACL}},
  year      = {2025},
  note      = {arXiv:2411.00348}
}

@article{debenedetti2025camel,
  title   = {Defeating Prompt Injections by Design},
  author  = {Debenedetti, Edoardo and Shumailov, Ilia and Fan, Tianqi and Hayes, Jamie and Carlini, Nicholas and Fabian, Daniel and Kern, Christoph and Shi, Chongyang and Terzis, Andreas and Tram{\`e}r, Florian},
  journal = {arXiv preprint arXiv:2503.18813},
  year    = {2025}
}

@article{zhang2026agentsentry,
  title   = {{AgentSentry}: Mitigating Indirect Prompt Injection in {LLM} Agents via Temporal Causal Diagnostics and Context Purification},
  author  = {Zhang, Tian and Xu, Yiwei and Wang, Juan and Guo, Keyan and Xu, Xiaoyang and Xiao, Bowen and Guan, Quanlong and Fan, Jinlin and Liu, Jiawei and Liu, Zhiquan and Hu, Hongxin},
  journal = {arXiv preprint arXiv:2602.22724},
  year    = {2026}
}

@article{he2026attriguard,
  title   = {{AttriGuard}: Defeating Indirect Prompt Injection in {LLM} Agents via Causal Attribution of Tool Invocations},
  author  = {He, Yu and Zhu, Haozhe and Li, Yiming and Shao, Shuo and Yao, Hongwei and Liu, Zhihao and Qin, Zhan},
  journal = {arXiv preprint arXiv:2603.10749},
  year    = {2026}
}

@article{lee2024infection,
  title   = {Prompt Infection: {LLM-to-LLM} Prompt Injection within Multi-Agent Systems},
  author  = {Lee, Donghyun and Tiwari, Mo},
  journal = {arXiv preprint arXiv:2410.07283},
  year    = {2024}
}

@article{jia2026mascope,
  title   = {Beyond Input Guardrails: Reconstructing Cross-Agent Semantic Flows for Execution-Aware Attack Detection},
  author  = {Wei, Yangyang and Xu, Yijie and Li, Zhenyuan and Shen, Xiangmin and Ji, Shouling},
  journal = {arXiv preprint arXiv:2603.04469},
  year    = {2026},
  note    = {NDSS 2026 poster}
}

@inproceedings{chen2024agentpoison,
  title     = {{AgentPoison}: Red-Teaming {LLM} Agents via Poisoning Memory or Knowledge Bases},
  author    = {Chen, Zhaorun and Xiang, Zhen and Xiao, Chaowei and Song, Dawn and Li, Bo},
  booktitle = {Advances in Neural Information Processing Systems ({NeurIPS})},
  year      = {2024}
}

@inproceedings{dong2025minja,
  title     = {A Practical Memory Injection Attack ({MINJA}) against {LLM} Agents},
  author    = {Dong, Shen and Xu, Shaochen and He, Pengfei and Li, Yige and Tang, Jiliang and Liu, Tianming and Liu, Hui and Xiang, Zhen},
  booktitle = {Advances in Neural Information Processing Systems ({NeurIPS})},
  year      = {2025},
  note      = {arXiv:2503.03704}
}

@article{zou2026etamp,
  title   = {Poison Once, Exploit Forever: Environment-Injected Memory Poisoning Attacks on Web Agents},
  author  = {Zou, Wei and Dong, Mingwen and Romero Calvo, Miguel and Chang, Shuaichen and Guo, Jiang and Lee, Dongkyu and Niu, Xing and Ma, Xiaofei and Qi, Yanjun and Jiang, Jiarong},
  journal = {arXiv preprint arXiv:2604.02623},
  year    = {2026}
}

@article{triedman2025hijack,
  title   = {Multi-Agent Systems Execute Arbitrary Malicious Code},
  author  = {Triedman, Harold and Jha, Rishi and Shmatikov, Vitaly},
  journal = {arXiv preprint arXiv:2503.12188},
  year    = {2025}
}

@inproceedings{shahroz2025agents,
  title     = {Agents Under Siege: Breaking Pragmatic Multi-Agent {LLM} Systems with Optimized Prompt Attacks},
  author    = {Shahroz, Rana and Tan, Zhen and Yun, Sukwon and Fleming, Charles and Chen, Tianlong},
  booktitle = {Proceedings of the 63rd Annual Meeting of the Association for Computational Linguistics (Volume 1: Long Papers)},
  pages     = {9661--9674},
  publisher = {Association for Computational Linguistics},
  doi       = {10.18653/v1/2025.acl-long.476},
  url       = {https://aclanthology.org/2025.acl-long.476/},
  year      = {2025}
}

@article{hossain2025pipeline,
  title   = {A Multi-Agent {LLM} Defense Pipeline against Prompt Injection Attacks},
  author  = {Hossain, S. M. Asif and Shayoni, Ruksat Khan and Ameen, Mohd Ruhul and Islam, Akif and Mridha, M. F. and Shin, Jungpil},
  journal = {arXiv preprint arXiv:2509.14285},
  year    = {2025}
}

@article{gosmar2025nlp,
  title   = {Prompt Injection Detection and Mitigation via {AI} Multi-Agent {NLP} Frameworks},
  author  = {Gosmar, Diego and Dahl, Deborah A. and Gosmar, Dario},
  journal = {arXiv preprint arXiv:2503.11517},
  year    = {2025}
}

@article{zheng2025agentsight,
  title   = {{AgentSight}: System-Level Observability for {AI} Agents Using {eBPF}},
  author  = {Zheng, Yusheng and Hu, Yanpeng and Yu, Tong and Quinn, Andi},
  journal = {arXiv preprint arXiv:2508.02736},
  year    = {2025}
}

@inproceedings{saha2024adapt,
  title     = {{ADAPT} It! Automating {APT} Campaign and Group Attribution by Leveraging and Linking Heterogeneous Files},
  author    = {Saha, Aakanksha and Blasco, Jorge and Cavallaro, Lorenzo and Lindorfer, Martina},
  booktitle = {27th International Symposium on Research in Attacks, Intrusions and Defenses ({RAID})},
  year      = {2024},
  doi       = {10.1145/3678890.3678909}
}

@inproceedings{cheng2024kairos,
  title     = {{KAIROS}: Practical Intrusion Detection and Investigation Using Whole-System Provenance},
  author    = {Cheng, Zijun and Lv, Qiujian and Liang, Jinyuan and Wang, Yan and Sun, Degang and Pasquier, Thomas and Han, Xueyuan},
  booktitle = {{IEEE} Symposium on Security and Privacy ({S\&P})},
  year      = {2024},
  note      = {arXiv:2308.05034}
}

@inproceedings{rehman2024flash,
  title     = {{FLASH}: A Comprehensive Approach to Intrusion Detection via Provenance Graph Representation Learning},
  author    = {Rehman, Mati Ur and Ahmadi, Hadi and Hassan, Wajih Ul},
  booktitle = {{IEEE} Symposium on Security and Privacy ({S\&P})},
  year      = {2024},
  note      = {{IEEE} Xplore document 10646725}
}

@article{wang2022threatrace,
  title   = {{THREATRACE}: Detecting and Tracing Host-Based Threats in Node Level through Provenance Graph Learning},
  author  = {Wang, Su and Wang, Zhiliang and Zhou, Tao and Sun, Hongbin and Yin, Xia and Han, Dongqi and Zhang, Han and Shi, Xingang and Yang, Jiahai},
  journal = {{IEEE} Transactions on Information Forensics and Security},
  volume  = {17},
  pages   = {3972--3987},
  year    = {2022}
}

@inproceedings{li2024nodlink,
  title     = {{NODLINK}: An Online System for Fine-Grained {APT} Attack Detection and Investigation},
  author    = {Li, Shaofei and Dong, Feng and Xiao, Xusheng and Wang, Haoyu and Shao, Fei and Chen, Jiedong and Guo, Yao and Chen, Xiangqun and Li, Ding},
  booktitle = {Network and Distributed System Security Symposium ({NDSS})},
  year      = {2024},
  note      = {arXiv:2311.02331}
}

@article{hassan2024hades,
  title   = {{HADES}: Detecting {Active Directory} Attacks via Whole Network Provenance Analytics},
  author  = {Liu, Qi and Bao, Kaibin and Hassan, Wajih Ul and Hagenmeyer, Veit},
  journal = {arXiv preprint arXiv:2407.18858},
  year    = {2024}
}

@inproceedings{venugopalan2025fp,
  title     = {{FP-Inconsistent}: Measurement and Analysis of Fingerprint Inconsistencies in Evasive Bot Traffic},
  author    = {Venugopalan, Hari and Munir, Shaoor and Ahmed, Shuaib and Wang, Tangbaihe and King, Samuel T. and Shafiq, Zubair},
  booktitle = {Proceedings of the {ACM} Internet Measurement Conference ({IMC})},
  year      = {2025},
  note      = {arXiv:2406.07647}
}

@inproceedings{go2025xdac,
  title     = {{XDAC}: {XAI}-Driven Detection and Attribution of {LLM}-Generated News Comments in {Korean}},
  author    = {Go, Wooyoung and Kim, Hyoungshick and Oh, Alice and Kim, Yongdae},
  booktitle = {Proceedings of the 63rd Annual Meeting of the Association for Computational Linguistics ({ACL})},
  pages     = {22728--22750},
  year      = {2025}
}

@article{stamatatos2009survey,
  title   = {A Survey of Modern Authorship Attribution Methods},
  author  = {Stamatatos, Efstathios},
  journal = {Journal of the American Society for Information Science and Technology},
  volume  = {60},
  number  = {3},
  pages   = {538--556},
  year    = {2009}
}

@article{hartigan1985dip,
  title   = {The Dip Test of Unimodality},
  author  = {Hartigan, J. A. and Hartigan, P. M.},
  journal = {The Annals of Statistics},
  volume  = {13},
  number  = {1},
  pages   = {70--84},
  year    = {1985}
}

@article{mcinnes2017hdbscan,
  title   = {hdbscan: Hierarchical Density Based Clustering},
  author  = {McInnes, Leland and Healy, John and Astels, Steve},
  journal = {Journal of Open Source Software},
  volume  = {2},
  number  = {11},
  pages   = {205},
  year    = {2017}
}

@inproceedings{rosenberg2007vmeasure,
  title     = {{V-Measure}: A Conditional Entropy-Based External Cluster Evaluation Measure},
  author    = {Rosenberg, Andrew and Hirschberg, Julia},
  booktitle = {Proceedings of the 2007 Joint Conference on Empirical Methods in Natural Language Processing and Computational Natural Language Learning ({EMNLP-CoNLL})},
  year      = {2007}
}

@misc{mitre2020,
  title        = {{MITRE} {ATT\&CK}: Enterprise Matrix},
  author       = {{The MITRE Corporation}},
  howpublished = {\url{https://attack.mitre.org/}},
  year         = {2020}
}

@inproceedings{debenedetti2024agentdojo,
  title     = {{AgentDojo}: A Dynamic Environment to Evaluate Attacks and Defenses for {LLM} Agents},
  author    = {Debenedetti, Edoardo and Zhang, Jie and Balunovi{\'c}, Mislav and Beurer-Kellner, Luca and Fischer, Marc and Tram{\`e}r, Florian},
  booktitle = {Advances in Neural Information Processing Systems ({NeurIPS})},
  year      = {2024}
}

@inproceedings{liu2024agentbench,
  title     = {{AgentBench}: Evaluating {LLMs} as Agents},
  author    = {Liu, Xiao and others},
  booktitle = {International Conference on Learning Representations ({ICLR})},
  year      = {2024}
}

@article{inan2023llamaguard,
  title   = {{Llama Guard}: {LLM}-Based Input-Output Safeguard for Human-{AI} Conversations},
  author  = {Inan, Hakan and Upasani, Kartikeya and Chi, Jianfeng and Rungta, Rashi and Iyer, Krithika and Mao, Yuning and Tontchev, Michael and Hu, Qing and Fuller, Brian and Testuggine, Davide and Khabsa, Madian},
  journal = {arXiv preprint arXiv:2312.06674},
  year    = {2023}
}

@article{jia2025critical,
  title   = {A Critical Evaluation of Defenses against Prompt Injection Attacks},
  author  = {Jia, Yuqi and Shao, Zedian and Liu, Yupei and Jia, Jinyuan and Song, Dawn and Gong, Neil Zhenqiang},
  journal = {arXiv preprint arXiv:2505.18333},
  year    = {2025}
}

@article{rabieinejad2026alteda,
  title   = {Beyond the Prompt: Log-Based Threat Detection and Attribution for Multi-Agent {LLMs}},
  author  = {Rabieinejad, Elnaz and Zarrinkalam, Fattane and Dehghantanha, Ali},
  journal = {Information Processing \& Management},
  volume  = {63},
  number  = {6},
  pages   = {104768},
  doi     = {10.1016/j.ipm.2026.104768},
  year    = {2026}
}

@misc{deepseekv42026,
  title        = {{DeepSeek-V4}: Towards Highly Efficient Million-Token Context Intelligence},
  author       = {{DeepSeek-AI}},
  howpublished = {Hugging Face model card for \texttt{deepseek-ai/DeepSeek-V4-Pro}, \url{https://huggingface.co/deepseek-ai/DeepSeek-V4-Pro}},
  note         = {Accessed via OpenRouter in 2026, model id \texttt{deepseek/deepseek-v4-pro}},
  year         = {2026}
}

@misc{qwen36_2026,
  title        = {{Qwen3.6-27B}: Model Card},
  author       = {{Qwen Team}},
  howpublished = {Hugging Face model card, \url{https://huggingface.co/Qwen/Qwen3.6-27B}},
  year         = {2026}
}

@misc{gpt55_2026,
  title        = {Introducing {GPT-5.5}},
  author       = {{OpenAI}},
  howpublished = {OpenAI release page, \url{https://openai.com/index/introducing-gpt-5-5/}},
  note         = {Accessed via OpenRouter in 2026, model id \texttt{openai/gpt-5.5-20260423}},
  year         = {2026}
}

@misc{gemma4_2026,
  title        = {{Gemma 4 31B IT}: Model Card},
  author       = {{Google DeepMind}},
  howpublished = {Hugging Face model card, \url{https://huggingface.co/google/gemma-4-31B-it}},
  year         = {2026}
}

@misc{opus47_2026,
  title        = {Introducing {Claude Opus 4.7}},
  author       = {{Anthropic}},
  howpublished = {Anthropic announcement, \url{https://www.anthropic.com/news/claude-opus-4-7}},
  note         = {Accessed via OpenRouter in 2026, model id \texttt{anthropic/claude-opus-4.7}},
  year         = {2026}
}

@misc{haiku45_2025,
  title        = {{Claude Haiku 4.5}: System Card},
  author       = {{Anthropic}},
  howpublished = {Anthropic system card, \url{https://assets.anthropic.com/m/99128ddd009bdcb/original/Claude-Haiku-4-5-System-Card.pdf}},
  year         = {2025}
}

@misc{gemini31_2026,
  title        = {{Gemini 3.1 Pro}: Model Card},
  author       = {{Google DeepMind}},
  howpublished = {Google DeepMind model card, \url{https://deepmind.google/models/model-cards/gemini-3-1-pro/}},
  note         = {Accessed via OpenRouter in 2026, model id \texttt{google/gemini-3.1-pro-preview}},
  year         = {2026}
}

@misc{llama4mav_2026,
  title        = {{Llama 4 Maverick 17B-128E}: Model Card},
  author       = {{Meta AI}},
  howpublished = {Hugging Face model card, \url{https://huggingface.co/meta-llama/Llama-4-Maverick-17B-128E}},
  note         = {Model release date April 5, 2025; accessed via OpenRouter in 2026, model id \texttt{meta-llama/llama-4-maverick}},
  year         = {2025}
}

@misc{llama33_2024,
  title        = {{Llama 3.3 70B Instruct}: Model Card},
  author       = {{Meta AI}},
  howpublished = {Hugging Face model card, \url{https://huggingface.co/meta-llama/Llama-3.3-70B-Instruct}},
  year         = {2024}
}

@article{azarafrooz2026cstm,
  title   = {Cross-Session Threats in {AI} Agents: Benchmark, Evaluation, and Algorithms},
  author  = {Azarafrooz, Ari},
  journal = {arXiv preprint arXiv:2604.21131},
  year    = {2026}
}

@inproceedings{gill2025binaryshield,
  title     = {{BinaryShield}: Cross-Service Threat Intelligence in {LLM} Services using Privacy-Preserving Fingerprints},
  author    = {Gill, Waris and Isak, Natalie and Dressman, Matthew},
  booktitle = {{IEEE} Conference on Secure and Trustworthy Machine Learning ({SaTML})},
  year      = {2026},
  note      = {arXiv:2509.05608}
}

@article{schroeder2025multiagentsec,
  title   = {Open Challenges in Multi-Agent Security: Towards Secure Systems of Interacting {AI} Agents},
  author  = {Schroeder de Witt, Christian and others},
  journal = {arXiv preprint arXiv:2505.02077},
  year    = {2025}
}

@article{dong2024agentops,
  title   = {{AgentOps}: Enabling Observability of {LLM} Agents},
  author  = {Dong, Liming and Lu, Qinghua and Zhu, Liming},
  journal = {arXiv preprint arXiv:2411.05285},
  year    = {2024}
}

@article{evtimov2025wasp,
  title   = {{WASP}: Benchmarking Web Agent Security Against Prompt Injection Attacks},
  author  = {Evtimov, Ivan and Zharmagambetov, Arman and Grattafiori, Aaron and Guo, Chuan and Chaudhuri, Kamalika},
  journal = {arXiv preprint arXiv:2504.18575},
  year    = {2025},
  url     = {https://arxiv.org/abs/2504.18575}
}

@inproceedings{li2025drift,
  title     = {{DRIFT}: Dynamic Rule-Based Defense with Injection Isolation for Securing {LLM} Agents},
  author    = {Li, Hao and Liu, Xiaogeng and Chiu, Hung-Chun and Li, Dianqi and Zhang, Ning and Xiao, Chaowei},
  booktitle = {Advances in Neural Information Processing Systems},
  year      = {2025},
  url       = {https://nips.cc/virtual/2025/poster/116028}
}

@inproceedings{luo2025agentauditor,
  title     = {{AgentAuditor}: Human-level Safety and Security Evaluation for {LLM} Agents},
  author    = {Luo, Hanjun and Dai, Shenyu and Ni, Chiming and Li, Xinfeng and Zhang, Guibin and Wang, Kun and Liu, Tongliang and Salam, Hanan},
  booktitle = {Advances in Neural Information Processing Systems},
  year      = {2025},
  url       = {https://nips.cc/virtual/2025/poster/120154}
}

@inproceedings{alperin2025masks,
  title     = {Masks and Mimicry: Strategic Obfuscation and Impersonation Attacks on Authorship Verification},
  author    = {Alperin, Kenneth and Leekha, Rohan and Uchendu, Adaku and Nguyen, Trang and Medarametla, Srilakshmi and Capote, Carlos Levya and Aycock, Seth and Dagli, Charlie},
  booktitle = {Proceedings of the 5th International Conference on Natural Language Processing for Digital Humanities},
  pages     = {102--116},
  url       = {https://aclanthology.org/2025.nlp4dh-1.10/},
  year      = {2025}
}
\bibliographystyle{icml2026}

\clearpage
\appendix
\onecolumn
\setcounter{table}{0}
\renewcommand{\thetable}{\Alph{table}}
\renewcommand{\theHtable}{appendix.\Alph{table}}
\setcounter{figure}{0}
\renewcommand{\thefigure}{\Alph{figure}}
\renewcommand{\theHfigure}{appendix.\Alph{figure}}

\section{Bootstrap Confidence Intervals}
\label{sec:bootstrap-ci}

We report fixed-similarity nonparametric bootstrap intervals for the main Full 2000 comparisons.
For V-measure, sessions are resampled with replacement from the evaluated scope and scored against the fixed clustering labels.
For pairwise AUC, campaign sessions are resampled with replacement and duplicate self-pairs are skipped.
These intervals quantify sample uncertainty for the reported similarity/cluster outputs; they do not rerun data generation, weight selection, or LLM-judge calls.
The same fixed-similarity session bootstrap (1{,}000 replicates, self-pairs skipped) is applied to the public DTap rows and the crossed-style hard slice by the released diagnostic scripts; those intervals are reported inline in \Cref{sec:supp-tables} and \Cref{sec:deployment-checklist}.

\begin{table}[!htbp]
\caption{Bootstrap 95\% confidence intervals for the main Full 2000 pairwise-AUC comparisons. Values are point estimate [2.5, 97.5] percentiles from 1{,}000 bootstrap replicates; the headline \aafv{} all-session V-measure CI is $0.269$ $[0.256,0.300]$.}
\label{tab:bootstrap-ci}
\vskip 0.1in
\centering
\setlength{\tabcolsep}{4pt}
\begin{tabular}{lc}
\toprule
Method & Pairwise AUC \\
\midrule
\aafv{}, Full 2000 & 0.818 [0.797, 0.837] \\
PromptShield score-delta & 0.521 [0.495, 0.548] \\
Peak+Accum score-delta & 0.522 [0.496, 0.550] \\
Best LLM judge (Opus 4.7) & 0.506 [0.501, 0.511] \\
\bottomrule
\end{tabular}
\vskip -0.1in
\end{table}

\FloatBarrier
\section{Residual-Channel Decomposition}
\label{sec:squeeze-appendix}

Let $c_d:\mathcal{T}_d\!\to\!\mathbb{R}_{\geq 0}$ denote the per-dimension residual-signal-reduction cost $c_d(\tau_d)=R_d(\mathrm{id},C)-R_d(\tau_d,C)$.
Under an approximate separability view, we use the following diagnostic accounting inequality for a joint transform $\tau=(\tau_{\mathrm{str}},\tau_{\mathrm{tmp}},\tau_{\mathrm{sty}})$ that drives $\sum_d w_d R_d(\tau,C)\le\epsilon$:
\begin{equation}
\label{eq:squeeze-lower}
\sum_d w_d c_d(\tau_d) \geq \sum_d w_d R_d(\mathrm{id},C)-\epsilon-\delta_{\mathrm{schema}}(\tau).
\end{equation}
Here $\delta_{\mathrm{schema}}(\tau)\geq0$ summarizes the schema-validity overhead required to keep the payload functional.
The decomposition is explanatory and diagnostic rather than a worst-case robustness theorem.

\FloatBarrier
\section{Supplementary Evaluation Tables}
\label{sec:supp-tables}

\begin{table}[!htbp]
\caption{Detailed campaign-linking results on Full 2000. Per-session detector rows are score-delta linkers; native pairwise linkers use standard cosine similarity over proxy-observed records without campaign labels at inference time; LLM-judge rows are chunked $40\times50$.}
\label{tab:baselines}
\vskip 0.1in
\centering
\setlength{\tabcolsep}{4pt}
\begin{tabular}{lcc}
\toprule
Method & V & AUC \\
\midrule
\multicolumn{3}{l}{\textit{Score-only detector adaptations}} \\
DataSentinel~\citep{liu2025datasentinel}     & 0.153 & 0.513 \\
PromptShield~\citep{jacob2025promptshield}   & 0.159 & 0.521 \\
Peak+Accum~\citep{hung2024attention}         & 0.146 & 0.522 \\
LlamaGuard-8B~\citep{inan2023llamaguard}     & 0.070 & 0.510 \\
\midrule
\multicolumn{3}{l}{\textit{Native pairwise linkers}} \\
Tool-sequence $n$-gram cosine                & -- & 0.784 \\
Prompt char/byte TF-IDF cosine               & -- & 0.707 \\
Unweighted structural+prompt cosine          & -- & \textbf{0.825} \\
\midrule
\multicolumn{3}{l}{\textit{Shared-runtime and LLM-judge adaptations}} \\
MAScope-async                               & 0.590 & 0.500 \\
GPT-5.5 judge~\citep{gpt55_2026}             & 0.155 & 0.505 \\
Claude Opus 4.7 judge~\citep{opus47_2026}    & 0.189 & 0.506 \\
Gemini 3.1 Pro judge~\citep{gemini31_2026}   & 0.169 & 0.504 \\
DeepSeek V4 Pro judge~\citep{deepseekv42026} & 0.173 & 0.502 \\
Llama 4 Maverick judge~\citep{llama4mav_2026}& 0.171 & 0.502 \\
\midrule
\textbf{\aafv{}}             & 0.269 & 0.818 \\
\bottomrule
\end{tabular}
\vskip -0.1in
\end{table}

\begin{table}[!htbp]
\caption{E2 ablation on Full 2000. S/T/Y = Structural / Temporal / prompt-text stylometry. Fixed subset rows use equal weights inside the subset; the final \aafv{} row uses the grid-selected weight.}
\label{tab:ablation}
\vskip 0.1in
\centering
\begin{tabular}{lcc}
\toprule
Subset & V-measure & Pairwise AUC \\
\midrule
S only & 0.230 & 0.784 \\
T only & 0.101 & 0.562 \\
Y only & 0.180 & 0.707 \\
S + T  & 0.179 & 0.698 \\
S + Y  & 0.292 & 0.825 \\
T + Y  & 0.130 & 0.619 \\
S + T + Y (equal) & 0.202 & 0.730 \\
\midrule
\textbf{\aafv{}} & \textbf{0.269} & \textbf{0.818} \\
\bottomrule
\end{tabular}
\vskip -0.1in
\end{table}

\begin{table}[!htbp]
\caption{E6 squeeze effect. Matched-persona triplet paired across evasion levels. Per-dimension columns report leave-one-in V-measure.}
\label{tab:evasion}
\vskip 0.1in
\centering
\begin{tabular}{lcccc}
\toprule
Level & S-only & T-only & Y-only & Full \\
\midrule
L0 (none)       & 0.240 & 0.100 & 0.180 & 0.295 \\
L1 (single dim) & 0.220 & 0.090 & 0.150 & 0.261 \\
L2 (triple dim) & 0.230 & 0.100 & 0.160 & 0.237 \\
L2 + CRF-light  & 0.250 & 0.100 & 0.160 & \textbf{0.260} \\
\bottomrule
\end{tabular}
\vskip -0.1in
\end{table}

\begin{table}[!htbp]
\caption{E1 sensitivity to agglomerative $K$ (Full 2000). Each row uses per-$K$ best $\mathbf{w}^*$ at grid resolution $0.1$.}
\label{tab:e1-ksweep}
\vskip 0.1in
\centering
\setlength{\tabcolsep}{4pt}
\begin{tabular}{lccc}
\toprule
$K$ & V-measure & Pairwise AUC & $\mathbf{w}^*$ (S/T/Y) \\
\midrule
10 & \textbf{0.335} & 0.825 & 0.5 / 0.0 / 0.5 \\
\textbf{15} & 0.319 & \textbf{0.828} & 0.4 / 0.0 / 0.6 \\
20 & 0.319 & \textbf{0.828} & 0.4 / 0.0 / 0.6 \\
\textbf{25} & 0.292 & 0.825 & 0.5 / 0.0 / 0.5 \\
30 & 0.266 & 0.809 & 0.7 / 0.0 / 0.3 \\
40 & 0.253 & 0.809 & 0.7 / 0.0 / 0.3 \\
50 & 0.248 & 0.818 & 0.6 / 0.0 / 0.4 \\
75 & 0.234 & 0.809 & 0.7 / 0.0 / 0.3 \\
\bottomrule
\end{tabular}
\vskip -0.1in
\end{table}

\begin{table}[!htbp]
\caption{E7 fixed persona-ID slice sensitivity on Full 2000. The reference slice uses IDs \texttt{p029}--\texttt{p040} and contains 11 campaign personas in this realization; it is not a schema-held-out campaign benchmark. Agglomerative clustering uses $K\!=\!25$ (persona count $+1$ noise label).}
\label{tab:persona-slice}
\vskip 0.1in
\centering
\begin{tabular}{lcc}
\toprule
Setting & V-measure & Pairwise AUC \\
\midrule
Other personas (13 campaigns)         & 0.187 & 0.615 \\
Reference slice (11 campaigns)        & 0.352 & 0.781 \\
\midrule
\textbf{All personas (E1, 24 campaigns)} & \textbf{0.269} & \textbf{0.818} \\
\bottomrule
\end{tabular}
\vskip -0.1in
\end{table}

\begin{table}[!htbp]
\caption{Leakage audit on the restored Full 2000 corpus. Verbatim hits are substring matches in prompts, tool arguments, and observations. Benign payload-family leakage is a gated check; adversarial payload-family hits are informational because the attack payload legitimately affects prompt content.}
\label{tab:leakage}
\vskip 0.1in
\centering
\setlength{\tabcolsep}{4pt}
\begin{tabular}{lccc}
\toprule
Label & Event hits & Session hits & Result \\
\midrule
persona\_id & 0/13{,}630 & 0/2{,}000 & PASS \\
payload\_family (benign) & 0/7{,}602 & 0/1{,}400 & PASS \\
payload\_family (adversarial) & 13/6{,}028 & 2/600 & INFO \\
style\_persona & 0/13{,}630 & 0/2{,}000 & PASS \\
evasion\_level / L0--L2 & 0/13{,}630 & 0/2{,}000 & PASS \\
tactic-internal terms & 0/13{,}630 & 0/2{,}000 & PASS \\
\bottomrule
\end{tabular}
\vskip -0.1in
\end{table}

\begin{table}[!htbp]
\caption{Role-slice sensitivity on Full 2000. Each row refits the \aafv{} grid on the stated subset. The \texttt{coder\_alt} slice is the schema-defined off-model coding role; these rows are controls against role/generator-slice confounding and are distinct from the independent generator regenerations reported in \Cref{tab:data-validation-diagnostics}.}
\label{tab:role-sensitivity}
\vskip 0.1in
\centering
\setlength{\tabcolsep}{4pt}
\begin{tabular}{lrrrr}
\toprule
Scope & Sessions & Campaign & V-measure & Pairwise AUC \\
\midrule
Full 2000 & 2000 & 300 & 0.269 & 0.818 \\
Without \texttt{coder\_alt} & 1690 & 252 & 0.291 & 0.818 \\
\texttt{coder\_alt} only & 310 & 48 & 0.435 & 0.745 \\
Non-coder roles & 1157 & 173 & 0.340 & 0.839 \\
Coder-family roles & 843 & 127 & 0.380 & 0.810 \\
\bottomrule
\end{tabular}
\vskip -0.1in
\end{table}

\begin{figure}[!htbp]
\centering
\includegraphics[width=\linewidth]{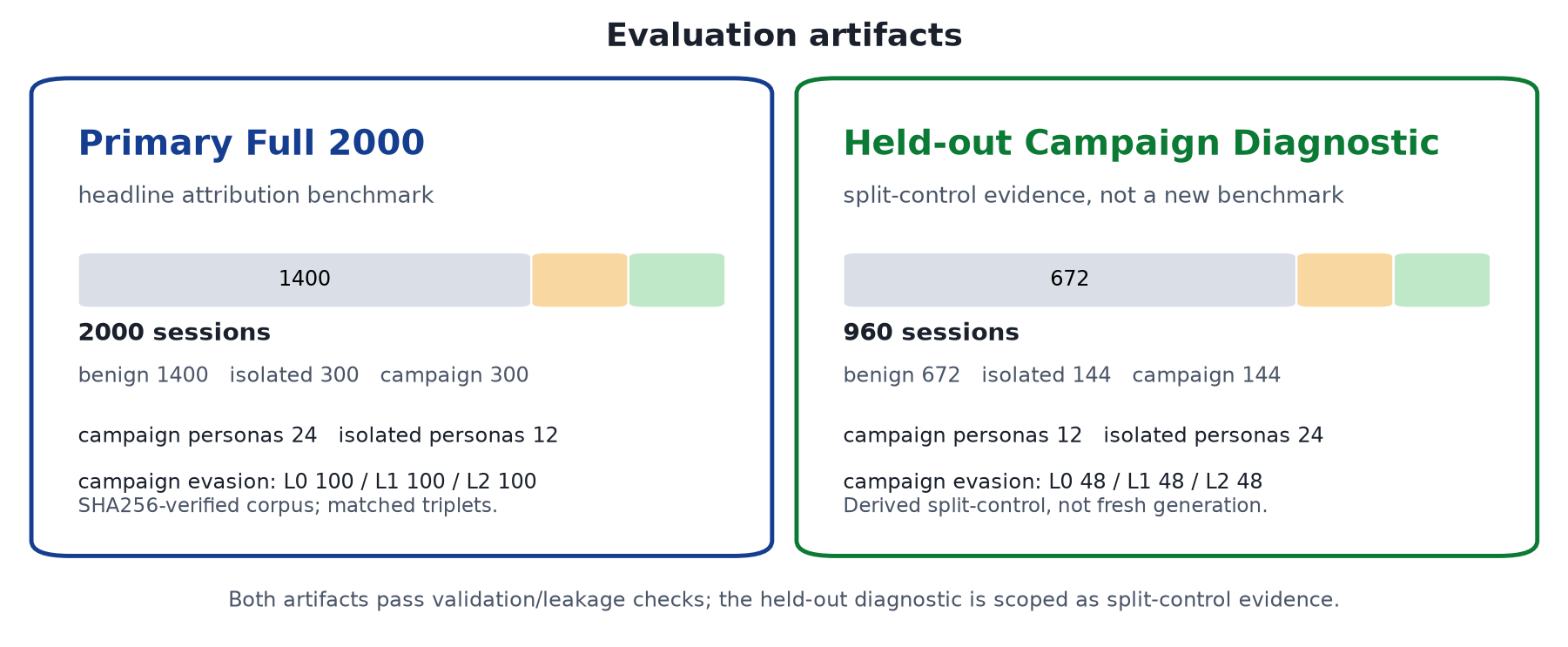}
\caption{Evaluation artifact map. Full 2000 is the primary headline benchmark; the held-out campaign diagnostic is a derived split-control artifact over 12 schema-held-out campaign personas, not an independently regenerated benchmark.}
\label{fig:data-map}
\vskip -0.1in
\end{figure}

\begin{figure}[!htbp]
\centering
\includegraphics[width=\linewidth]{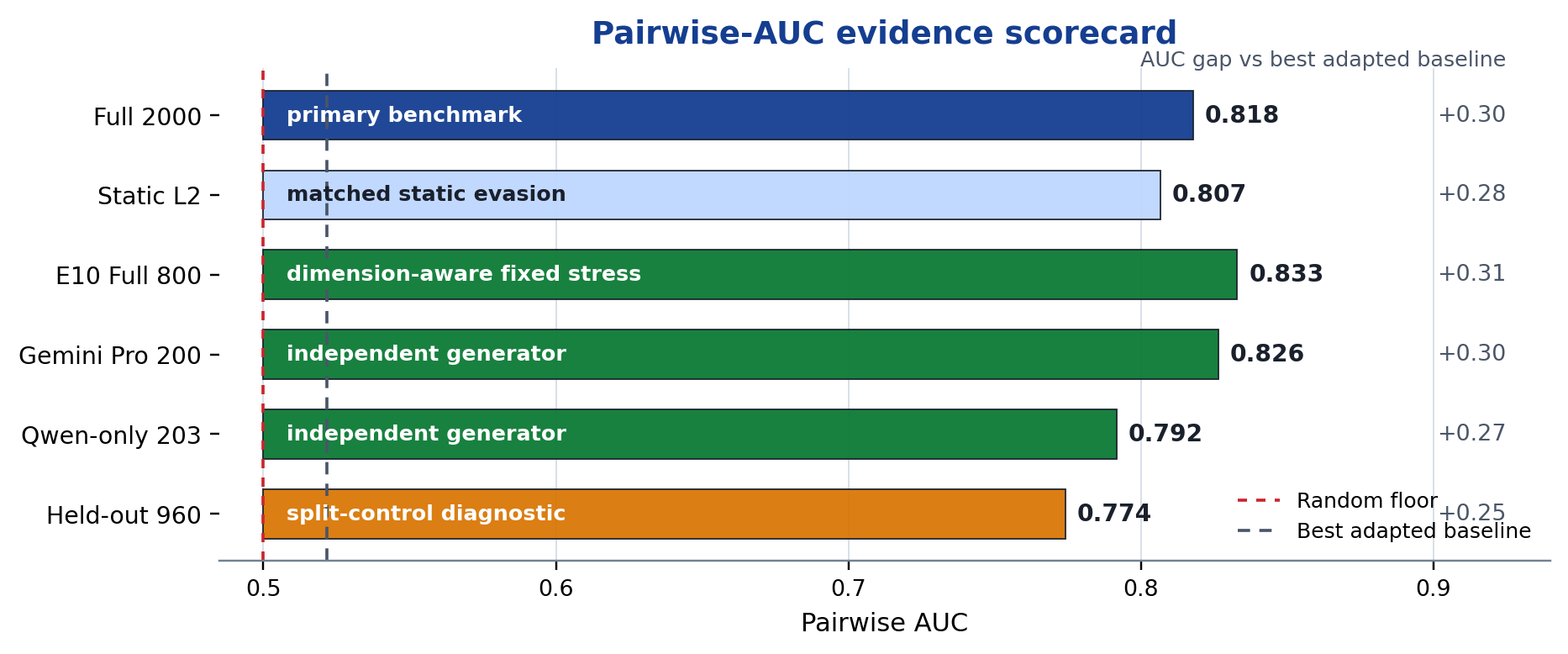}
\caption{Pairwise-AUC evidence scorecard across same-metric evidence axes. Bars show \aafv{}'s margin above the random-linking floor; dashed references mark random linking and the best adapted baseline. Rows mix primary, static-evasion, independent-generator, held-out, and E10 controls, so the figure summarizes evidence breadth rather than a single benchmark.}
\label{fig:robustness-scorecard}
\vskip -0.1in
\end{figure}

\begin{table}[!htbp]
\caption{Follow-up data-validation diagnostics on the released artifacts. These are control and deployment-style checks, not additional headline claims; the archive is generated by the packaged reproduction scripts.}
\label{tab:data-validation-diagnostics}
\vskip 0.1in
\centering
\setlength{\tabcolsep}{4pt}
\begin{tabular}{>{\raggedright\arraybackslash}p{0.30\textwidth}
                >{\raggedright\arraybackslash}p{0.24\textwidth}
                >{\raggedright\arraybackslash}p{0.38\textwidth}}
\toprule
Diagnostic & Value & Interpretation \\
\midrule
All-pair Precision@20 & 0.100 (2/20) & Harsh alert-budget view over 1{,}999{,}000 candidate pairs; useful as a ranking sanity check, not a finished deployment interface. \\
All-pair Recall@100 & 0.003 (5/1{,}728) & Top-100 pair alerts recover only a small fraction of all positives; operational deployments should filter or window candidates before attribution. \\
Within-payload-family AUC & 0.599 & Same payload family alone does not explain the 0.818 campaign-pair AUC. \\
Schema-held-out split control & V 0.293; AUC 0.774 & Split-control artifact over 12 schema-held-out personas, constructed from generated held-out isolated rows with matched L0/L1/L2 transforms. \\
Independent generator regenerations & Qwen 203: V 0.205, AUC 0.792; Gemini Pro 200: V 0.261, AUC 0.826 & Qwen-only and Gemini-Pro-only held-out campaign slices with 11/11 and 10/10 matched triplets, validation/leakage PASS, and fixed Full 2000 $\mathbf{w}^*=(0.6,0,0.4)$ rather than slice-specific tuning. \\
Distributional audit from stylometry & persona 0.400; payload 0.906; style 0.928 & Surface text can recover generated style attributes, so SCD-v1 should not be described as distributionally artifact-free. \\
Schema-validity sanity & L0/L1/L2 payload-valid 0.740/0.770/0.770 & Executable success-predicate check only; not a complete semantic proof of task validity. \\
Fixed-weight static evasion & AUC in [0.788, 0.828] & Holding $\mathbf{w}^*=(0.6,0,0.4)$ avoids per-row weight tuning and preserves above-random signal across packaged variants. \\
Fixed-weight E10 mirror & 0.852 $\to$ 0.833 & Full 800 diagnostic under the same fixed $\mathbf{w}^*$; separate from the released-script best-weight E10 diagnostic in \Cref{sec:e10-reproduction-gap}. \\
\bottomrule
\end{tabular}
\vskip -0.1in
\end{table}

\begin{table}[!htbp]
\caption{External public-corpus and crossed-style diagnostics. These rows separate public trajectory evidence and style-control probes from the compact released-artifact checks in \Cref{tab:data-validation-diagnostics}.}
\label{tab:external-style-diagnostics}
\vskip 0.1in
\centering
\setlength{\tabcolsep}{4pt}
\begin{tabular}{>{\raggedright\arraybackslash}p{0.30\textwidth}
                >{\raggedright\arraybackslash}p{0.24\textwidth}
                >{\raggedright\arraybackslash}p{0.38\textwidth}}
\toprule
Diagnostic & Value & Interpretation \\
\midrule
Public DTap feature-only diagnostic & V 0.362; AUC 0.734 [0.660, 0.810] & External public trajectory check over 160 feature-only adapted DTap sessions. Labels are path-derived campaign proxies, not true attacker identities, and raw prompts/tool parameters are not persisted; with the stylometric block empty under redaction, the fixed score reduces to structural evidence. \\
Public DTap include-text diagnostic & V 0.466; AUC 0.783 [0.711, 0.856]; same-domain AUC 0.713 [0.640, 0.788] & Include-text re-evaluation of the same 160 public sessions (Apache-2.0 source) restores the stylometric block (stylometry-only 0.719). Six of eight proxy groups are domain-pure, and restricting to same-domain pairs keeps the fixed score above random, so shared task domain alone does not explain the signal. \\
Fresh crossed-style Qwen diagnostic & \aafv{} all/hard AUC 0.747/0.599 [0.538, 0.662]; structural-only hard AUC 0.686 [0.621, 0.744]; stylometry-only hard AUC 0.271 [0.184, 0.367] & Fresh 120-session Qwen generation: each attacker appears under two style directives, and each style directive is shared by multiple attackers. Brackets are 1{,}000-replicate session-bootstrap 95\% CIs on the hard slice. The hard slice is above random for structural residue but shows that the fixed full score remains style-sensitive. \\
Fresh crossed-style Gemini regeneration & \aafv{} all/hard AUC 0.747/0.572 [0.502, 0.633]; structural-only hard AUC 0.686 [0.625, 0.739]; stylometry-only hard AUC 0.197 [0.123, 0.286] & Independent Gemini-3.1-Pro regeneration of the same 120-session crossed-style protocol with zero failed generation events. The structural hard-slice point matches the Qwen slice, so structural residue surviving style crossing is not a single-generator artifact. \\
\bottomrule
\end{tabular}
\vskip -0.1in
\end{table}

\begin{table}[!htbp]
\caption{Fresh crossed-style Qwen diagnostic. The hard slice uses same-attacker/different-style pairs as positives and different-attacker/same-style pairs as negatives. This control tests whether linking collapses when generated style is deliberately crossed with attacker identity in fresh model-generated sessions.}
\label{tab:crossed-style}
\vskip 0.1in
\centering
\begin{tabular}{lcc}
\toprule
Method & All crossed AUC & Hard-slice AUC \\
\midrule
Stylometry-only & 0.721 & 0.271 \\
Structural-only & 0.686 & \textbf{0.686} \\
Temporal-only & 0.602 & 0.582 \\
Unweighted structural+prompt & \textbf{0.761} & 0.554 \\
\textbf{\aafv{} fixed} & 0.747 & 0.599 \\
\bottomrule
\end{tabular}
\vskip -0.1in
\end{table}

\FloatBarrier
\section{Telemetry Degradation Diagnostic}
\label{sec:telemetry-degradation}

To evaluate sensitivity to incomplete or privacy-minimized proxy logs, we add a no-API degradation diagnostic over the released Full 2000 corpus and both native-framework probes.
The transform is deterministic except for event dropout: we coarsen or zero timestamps, redact prompts, redact tool arguments and responses, and randomly remove tool events with five fixed seeds while keeping at least one event per session.
\Cref{tab:telemetry-degradation} reports fixed-weight pairwise AUC; these rows are stress tests over existing artifacts, not production telemetry or new headline benchmarks.
Timestamp coarsening and missing timestamps leave the fixed score unchanged because the deployed weight is $\mathbf{w}^*=(0.6,0,0.4)$, which already zeros temporal evidence.

\begin{table}[!htbp]
\caption{No-API telemetry degradation diagnostic. Dropout rows report mean [min, max] fixed-weight pairwise AUC across five deterministic seeds. Base rows recompute fixed-weight AUC over campaign pairs with the per-scope structural adapter, so they can differ from \Cref{tab:bridge-checks} in the final digit.}
\label{tab:telemetry-degradation}
\vskip 0.1in
\centering
\setlength{\tabcolsep}{4pt}
\begin{tabular}{lccccc}
\toprule
Scope & Base & Prompt redacted & Args/resp. redacted & 30\% event drop & 50\% event drop \\
\midrule
Full 2000 & 0.818 & 0.784 & 0.819 & 0.792 [0.788, 0.795] & 0.757 [0.751, 0.764] \\
LangGraph native & 0.664 & 0.564 & 0.666 & 0.652 [0.631, 0.684] & 0.647 [0.634, 0.665] \\
OpenClaw native & 0.704 & 0.581 & 0.653 & 0.704 [0.699, 0.710] & 0.701 [0.698, 0.705] \\
\bottomrule
\end{tabular}
\vskip -0.1in
\end{table}

The diagnostic sharpens, rather than removes, the deployment limitation.
Event dropout causes a gradual margin reduction on Full 2000 and a smaller change on the compact native probes, while prompt redaction is more damaging on native traces, where stylometric and framework response residue carry more of the fixed-weight signal.
Argument/response redaction barely changes Full 2000 because the primary structural contract is tool-sequence based, but it weakens OpenClaw because the native-resource structural adapter uses coarse argument resource classes.
Thus the operational recommendation is to preserve tool-call order and minimal prompt or session-summary evidence when governance allows, and to report logging completeness before comparing field traces with SCD-v1.
Additional native-runtime replay artifacts are included in the supplement only as implementation and attenuation notes; they are not used as main-paper evidence.

\FloatBarrier
\section{Case Study: \aafv{} on Native OpenClaw Traces}
\label{sec:openclaw-case-study}

We complement the synthetic SCD-v1 evaluation with a controlled OpenClaw native-framework trace probe.
This appendix is a schema-compatibility and deployment-plausibility check, not a benchmark.
The source execution uses a local OpenClaw agent profile in a sandboxed enterprise-style helpdesk setting; OpenClaw delegates to Claude Opus 4.7~\citep{opus47_2026} and Claude Haiku 4.5~\citep{haiku45_2025} through its local CLI path.
We release only sanitized SCD-v1-style derived records; raw local OpenClaw session logs and Claude CLI project logs are excluded.

\paragraph{Setup.}
We script 24 F1 probes per backend: 12 adversarial-shaped campaign tickets and 12 benign routine-ops tickets against sandboxed dummy credential, SSH-key, and authentication-log files.
Campaign tickets cover credential-pretext, inventory-exfiltration, and handover-orientation families; benign tickets cover routine file existence, listing, count, and log-summary requests.
All sessions use one controlled OpenClaw helpdesk profile, so this is a single-framework/single-role native trace probe rather than a multi-agent routing evaluation.

\paragraph{Sanitized artifact.}
The released artifact contains 48 derived native executions and 98 tool events: 24 Opus executions with 52 tool events and 24 Haiku executions with 46 tool events.
Sanitization rewrites local paths, removes CLI binding paths from metadata, and replaces credential-shaped sandbox strings with inert placeholders.
The OpenClaw leakage report finds 0 remaining hits for local user paths, local user tokens, prior-track tokens, CLI state paths, and credential-shaped strings.
The raw native trace is not included in the public supplement; only the sanitized derived JSONLs and metric summary are packaged.

\paragraph{\aafv{} result.} Holding the Full 2000 fixed deployment weight $\mathbf{w}^*=(0.6,0,0.4)$, we evaluate the native traces with a resource-aware structural view that canonicalizes wrapper status, operation class, and coarse resource class from tool arguments rather than relying only on raw tool names.
This adapter is used only for the OpenClaw native-trace probe; all SCD-v1 headline results retain the original tool-sequence structural feature contract.
\aafv{} reaches V/AUC $0.231/0.704$ on the combined 48-execution OpenClaw probe.
The Haiku slice reaches V/AUC $0.318/0.722$, and the Opus slice improves to $0.317/0.639$; the archived JSON also reports the legacy tool-only diagnostic, where Opus was the weakest slice.
Stylometry-only remains informative on both backends (AUC $0.714$ Opus, $0.717$ Haiku), while native structural and temporal signals are backend-sensitive.
We therefore use the result as evidence that the pipeline runs on native framework traffic and that fixed-weight signal survives across both backend slices, not as a headline performance estimate.

\paragraph{Implication.}
The OpenClaw probe illustrates three deployment-plausibility points for \S\ref{sec:threat}: native LLM-agent framework traces can (i)~contain tool-use and refusal patterns not present in synthetic harness records, (ii)~be mapped into the same proxy-observed schema consumed by \aafv{}, and (iii)~expose backend-specific failure boundaries that a synthetic-only benchmark would hide.
The Opus/Haiku split is especially useful as a caution: adding real native executions can strengthen acceptance evidence only when the paper keeps the scope narrow and reports boundary cases rather than smoothing them away.

\paragraph{Caveats.}
This OpenClaw probe is controlled local execution, not production telemetry, field prevalence, or a real-world benchmark.
It uses one sandboxed tenant, one OpenClaw helpdesk profile, and two Claude-family backend slices; broader cross-framework, cross-role, and real-trace validation remain future work.
The primary quantitative claims remain supported by the SCD-v1 artifacts; the OpenClaw row and appendix should be read as separate-scope deployment-plausibility evidence for native trace compatibility.

\FloatBarrier
\section{Second Native-Framework Probe: LangGraph Traces}
\label{sec:langgraph-case-study}

To check that the SCD-v1 derived schema is not specific to one native agent framework, we add a second controlled probe using LangGraph (langgraph 1.1.10, langchain 1.2.17).
The setup mirrors \S\ref{sec:openclaw-case-study}: a single-agent ReAct helpdesk profile inside a sandboxed enterprise-style tenant under \texttt{sandbox/langgraph\_native/}, with dummy credential, SSH-key, and authentication-log files.
All probed paths point to dummy files; no real secrets, private repository paths, or production telemetry are involved.
This is single-agent native trace evidence for schema portability; campaign correlation remains asynchronous cross-deployment as defined in \S\ref{sec:threat}.

\paragraph{Setup and sanitized artifact.}
We script 120 native executions across two backend slices served via OpenRouter with reasoning disabled, Qwen3.6-27B~\citep{qwen36_2026} and Llama-3.3-70B-Instruct~\citep{llama33_2024}: 60 adversarial-shaped campaign tickets and 60 benign routine-ops tickets, balanced across five linguistically distinct personas (audit-tight, compliance-verbose, casual-ops, log-forensics, config-archaeology) with deliberately divergent tool-use habits and multi-step prompts.
The released artifact contains 120 derived sessions and 600 events (mean 5.77 events per session for the Qwen slice and 4.23 for the Llama slice).
Sanitization rewrites local paths, normalizes framework cache paths, drops captured API key fields from \texttt{\_meta}, and replaces credential-shaped sandbox strings with inert placeholders.
The LangGraph leakage report finds 0 remaining hits across nine local-or-secret patterns on both backend slices; an independent grep for absolute user paths, AKIA tokens, OpenAI/Anthropic key shapes, and operator names also returns 0.
Raw LangGraph trace files and framework cache directories are excluded from the public supplement.
The supplement also includes a 480-execution multi-role stress variant across four role-specific dummy sandboxes (\texttt{ops\_helpdesk}, \texttt{code\_review\_agent}, \texttt{finance\_ops\_agent}, and \texttt{data\_analyst\_agent}), five persona voices, and the same Qwen/Llama backend slices.
This artifact contains 240 Qwen sessions and 240 Llama sessions over 2{,}845 events; its sanitization report reduces 370 raw local-or-secret hits to 0 remaining hits.
One Llama benign execution hits a LangGraph recursion limit and is retained as a sanitized error record; removing it changes combined AUC only from $0.71252$ to $0.71259$.

\paragraph{\aafv{} result.} Holding the same Full 2000 fixed deployment weight $\mathbf{w}^*=(0.6,0,0.4)$ and the resource-aware structural view used for OpenClaw, \aafv{} reaches combined V/AUC $0.221/0.663$ on the 120-execution LangGraph probe (330 positive pairs over 1770 campaign pairs); the Qwen slice reaches $0.235/0.650$ and the Llama slice reaches $0.190/0.621$.
Stylometry-only is the most informative single dimension at combined AUC $0.774$, the structural-stylometry equal-weight setting reaches $0.687$, and structural-only at $0.564$ stays above random while remaining the weakest dimension on this trace; temporal-only is near random ($0.515$), consistent with the deployment weight that already zeros temporal contribution.
Unlike the OpenClaw probe, the legacy tool-only structural diagnostic actually rises here to combined V/AUC $0.261/0.692$, slightly above the resource-aware view.
We retain the resource-aware mode as the cross-probe default for consistency with \S\ref{sec:openclaw-case-study} but report both diagnostics.
On the 480-execution multi-role stress variant, the same fixed weights reach combined V/AUC $0.260/0.713$ under the resource-aware structural view (Qwen $0.290/0.727$, Llama $0.249/0.706$).
Stylometry-only is again the strongest single channel (combined AUC $0.841$), structural-only is weaker under the resource-aware view (AUC $0.538$), and the archived legacy tool-only diagnostic reaches AUC $0.802$.
We therefore treat the multi-role result as a stress-test boundary: the proxy schema and fixed-weight signal survive a larger cross-role native trace, but the signal is still partly carried by prompt-style and tool-habit residue and should not be read as field performance.

\paragraph{Implication.}
The LangGraph probes extend the OpenClaw observations to a different native framework runtime and a different model family: the SCD-v1 derived schema absorbs LangGraph state-machine events (intake, agent, tools, respond) without changing the headline feature contract; fixed-weight signal survives backend, framework, and controlled role changes; and stylometry remains the most transferable dimension across both probes.
The cross-trace divergence between resource-aware and tool-only structural views is itself informative: it shows that the structural adapter carries different weight in different native runtimes, and that the headline SCD-v1 results, which use neither, do not depend on the choice.
The native-resource structural mode is therefore strictly a native-trace adapter and does not enter any SCD-v1 headline number.

\paragraph{Caveats.}
These LangGraph probes are controlled local execution, not production telemetry, field prevalence, or a real-world benchmark.
The single-role probe uses one sandboxed tenant and one LangGraph helpdesk profile; the multi-role variant adds controlled role-specific dummy sandboxes but still does not represent real deployment traffic.
Additional native frameworks, adversary-aware tool sequences, and privacy-reviewed real traces remain future work.
The probes are intended as native-framework schema-compatibility and deployment-plausibility evidence; primary quantitative claims continue to rest on the SCD-v1 artifacts.

\FloatBarrier
\section{\aafv{} Qualitative Outcomes and Failure Modes}
\label{sec:qual-failures}

\Cref{tab:qual-failures} summarizes the qualitative readout behind \aafv{}'s aggregate metrics.
These cases are not additional benchmark points; they spell out when \aafv{} has enough proxy residue to carry attacker identity and when the evidence weakens enough that an operator should lower confidence or hand the cluster to a verifier.

\begin{table}[!htbp]
\caption{Representative \aafv{} outcomes and failure boundaries. All evidence is drawn from the reported OpenClaw case study, Full 2000 evaluation, or Full 800 non-oracle stress test.}
\label{tab:qual-failures}
\vskip 0.1in
\centering
\setlength{\tabcolsep}{4pt}
\renewcommand{\arraystretch}{1.08}
\begin{tabular}{>{\raggedright\arraybackslash}p{0.17\textwidth}
                >{\raggedright\arraybackslash}p{0.30\textwidth}
                >{\raggedright\arraybackslash}p{0.25\textwidth}
                >{\raggedright\arraybackslash}p{0.20\textwidth}}
\toprule
Outcome / failure mode & Observed evidence & \aafv{} interpretation & Operator response \\
\midrule
Success: proxy residue links a campaign &
\texttt{P\_credential\_pretext}, \texttt{P\_inventory\_exfil}, and \texttt{P\_handover\_orient} tickets remain above random on the combined 48-execution OpenClaw probe, with fixed-weight AUC $0.704$ overall. &
Native tool-use traces can be mapped into the same proxy schema, and resource-aware structural features preserve usable signal across both backend slices. &
High-confidence correlation requires observable proxy residue; preserve tool-call order and timing logs. \\
\midrule
Failure: backend-specific native residue &
The legacy tool-only structural diagnostic left Opus near random, while the native-resource structural view raises Opus fixed-weight AUC to $0.639$. &
Native framework traces can be dominated by backend-specific tool policy, refusal style, or repeated tool-choice skeletons unless structural features encode coarse operation and resource residue. &
Treat native-trace evidence as deployment-plausibility and inspect both backend slices and structural feature modes before claiming transfer. \\
\midrule
Failure: benign behavior also clusters &
The OpenClaw probe includes 24 benign helpdesk sessions, and the Full 2000 all-session V-measure is modest even while campaign-pair AUC remains high. &
\aafv{} links recurring behavior; it is not a benign-vs-attack classifier and can cluster benign teams by stable work style. &
Use \aafv{} after first-stage detection or logging filters; do not turn clusters directly into block decisions. \\
\midrule
Boundary: dimension-aware pressure reduces margin &
With the fixed Full 2000 weight, the Full 800 E10 mirror lowers AUC from $0.852$ to $0.833$; the released-script diagnostic reaches $0.843$ after shifting weights toward prompt stylometry $(0.4,0.0,0.6)$, and the archived seven-source probe reaches $0.808$ at seed 20260505. &
Attacking multiple dimensions shrinks the margin in released artifacts, but surviving residues remain enough to stay above random without detector-score feedback. &
Fully-adaptive L3 with oracle remains out of scope; monitor confidence and refresh weights as attacks adapt. \\
\midrule
Boundary: small-$n$ LLM judge can win &
On small native traces, an LLM verifier can inspect all sessions in one context; at Full 2000, chunked LLM judges fall to AUC $0.500$--$0.506$ under the linking protocol. &
Semantic inspection may beat proxy fingerprints when all sessions fit in one context, but chunking breaks global identity at scale. &
Use \aafv{} as the scalable first pass, then spend LLM calls on the $K$ discovered clusters. \\
\bottomrule
\end{tabular}
\renewcommand{\arraystretch}{1.0}
\vskip -0.1in
\end{table}

The main \aafv{} failure pattern is therefore not a single catastrophic miss, but a confidence drop under missing or deliberately weakened residue.
Refusals and chat-only sessions remove structural evidence; benign teams with stable work style can form non-attack clusters; and dimension-aware non-oracle compositions can reduce the separation margin.
In all three cases, the appropriate operational response is to treat \aafv{} as a prioritization layer: retain the candidate links, expose the contributing dimensions, and escalate low-margin clusters to analyst or per-cluster LLM review.

\FloatBarrier
\section{Operational Validity Criteria Beyond SCD-v1}
\label{sec:deployment-checklist}

This appendix summarizes operational validity criteria for interpreting \aafv{} beyond SCD-v1.
The criteria are phrased as acceptance criteria rather than as additional evidence: the current paper provides a controlled benchmark and native-framework probes, while a real trace study would need independent campaign ground truth and privacy review.

\begin{table}[!htbp]
\caption{Operational validity criteria beyond SCD-v1. The middle columns identify what can degrade in production and what a future field study should report before claiming external validity.}
\label{tab:field-checklist}
\vskip 0.1in
\centering
\footnotesize
\setlength{\tabcolsep}{3pt}
\renewcommand{\arraystretch}{1.0}
\begin{tabular}{>{\raggedright\arraybackslash}p{0.17\textwidth}
                >{\raggedright\arraybackslash}p{0.26\textwidth}
                >{\raggedright\arraybackslash}p{0.30\textwidth}
                >{\raggedright\arraybackslash}p{0.19\textwidth}}
\toprule
Issue & Production failure mode & Validation criterion & Operational stance \\
\midrule
Synthetic benchmark dependence &
SCD-v1 controls persona, payload family, and evasion transforms; real traffic may contain tool catalogs, benign tasks, and attack behaviors outside the harness. &
Report a time-split trace with representative benign teams, tool families, and agent roles; publish the trace schema and sampling policy even if raw prompts require gated access. &
Treat SCD-v1 as a reproducible protocol and stress test, not as prevalence or field-performance evidence. \\
\midrule
Native-probe gap &
OpenClaw and LangGraph AUCs are lower than the primary result, and backend/tool-wrapper policies shift which structural features matter. &
Run each framework/backend slice separately; report both raw tool-name and resource-aware structural views; explain any slice where signal collapses. &
Use native probes as schema-portability checks, not headline benchmarks. \\
\midrule
Style confounding &
The stylometric audit recovers generated style attributes, and the hard crossed-style slice lowers fixed-weight \aafv{} from the Full 2000 headline to AUC $0.599$. &
Establish campaign identity independently of prompt style, for example through incident-response ground truth, controlled red-team operators, or infrastructure indicators. &
Expose structural, stylometric, and temporal contributions separately and avoid style-only escalation. \\
\midrule
Temporal block weakness &
The deployed weight zeros temporal evidence, and temporal-only is near random in the reported native traces. &
Only claim timing transfer when proxy logs contain reliable queue delays, retries, fallback behavior, and cross-session interleaving not injected by a harness. &
Keep timing as a diagnostic channel and do not describe it as an independent primary signal here. \\
\midrule
Benign team clusters &
Stable developer or team habits can form dense non-attack clusters and depress all-session partition quality. &
Evaluate post-filter traffic and report benign-only cluster audits, false-cluster examples, and analyst workload under a fixed alert budget. &
Use \aafv{} after first-stage detection or risk filtering; never convert clusters directly into blocks. \\
\midrule
Compute and storage &
All-pair scoring is quadratic inside a window, and raw prompt retention increases privacy and storage risk. &
Report window size, sharding rule, candidate-pair count, top-$K$ escalation budget, and whether fingerprints or raw prompts are stored. &
Shard by tenant, time, tool family, or detector pre-filter; store sparse fingerprints and pointers where possible. \\
\midrule
Privacy and governance &
Stylometric and structural fingerprints can encode personal writing habits, team workflows, or sensitive tool-use patterns. &
Document retention periods, access controls, cross-tenant restrictions, redaction policy, and whether secure aggregation or private set intersection is required. &
Prefer administrative-domain-local correlation; require explicit governance for cross-tenant matching. \\
\midrule
Adaptive L3 adversaries &
An attacker with detector-score feedback could optimize against the fixed score and reduce residual margins beyond the static and non-oracle stress tests. &
Separate non-oracle evasion from oracle feedback experiments; report score-query access, adaptive budget, and confidence-margin degradation over time. &
Monitor margins, refresh calibration, and route low-margin clusters to analyst or LLM review rather than claiming worst-case robustness. \\
\bottomrule
\end{tabular}
\renewcommand{\arraystretch}{1.0}
\vskip -0.1in
\end{table}

\paragraph{Style-crossing interpretation.}
The crossed-style diagnostic is the strongest current check against a pure style-artifact explanation, but it is not a style-invariant attribution benchmark.
On the hard slice, structural-only AUC reaches $0.686$ (session-bootstrap 95\% CI $[0.621,0.744]$), stylometry-only collapses to $0.271$ $[0.184,0.367]$, and the fixed full score reaches $0.599$ $[0.538,0.662]$ because its global weight still allocates mass to stylometric residue.
A deterministic per-persona variance summary gives structural-only hard-slice mean $0.688$ with range $[0.454,0.873]$ and one of twelve personas below random; the fixed full score has mean $0.601$ with range $[0.325,0.775]$ and two of twelve below random.
An independent Gemini-3.1-Pro regeneration of the same crossed-style protocol (120 sessions, matched persona/style design, zero failed generation events) reproduces the pattern: structural-only hard-slice AUC $0.686$ $[0.625,0.739]$, stylometry-only $0.197$ $[0.123,0.286]$, fixed score $0.572$ $[0.502,0.633]$, with per-persona structural mean $0.687$ and one of twelve personas below random.
The agreement of the structural channel across two generator families argues that structural residue surviving style crossing is not a single-generator artifact.
Thus the conservative reading is that structural residue usually survives style crossing in a small fresh slice, but the effect varies by persona and the headline Full 2000 signal remains partly style-correlated.
Real deployments should expose structural/stylometric contributions separately and reweight or audit before use on human operators.

\paragraph{Timing interpretation.}
The residual-channel hypothesis includes timing because real proxies may observe queueing, fallback retries, tool timeout behavior, and cross-session interleaving that are difficult for a non-oracle attacker to coordinate.
The present evidence does not show that this channel is strong: SCD-v1 chooses a zero temporal deployment weight, and the LangGraph temporal-only diagnostic is near random.
Timing should therefore be logged and reported as part of the feature contract, but future work must earn any claim that it is an independently useful deployment signal.

\paragraph{Operational triage path.}
At deployment time, \aafv{} should sit between a first-stage detector/logging filter and expensive review.
For a window with $m$ candidate sessions, the proxy computes sparse fingerprints, shards candidate pairs, ranks links by $K_{\mathbf{w}}$, exposes feature contributions and representative sessions, and escalates only high-margin or analyst-budgeted clusters.
This workflow is the intended interpretation of the alert-budget diagnostics: low raw Recall@100 on all candidate pairs is acceptable only if the production system first narrows the candidate window and treats \aafv{} as prioritization rather than autonomous blocking.

\FloatBarrier
\section{E10 Reproduction Gap}
\label{sec:e10-reproduction-gap}

The Full 800 dimension-aware non-oracle row in \Cref{tab:bridge-checks} reports the fixed-weight mirror archived with the released reproduction artifacts.
Under the same fixed $\mathbf{w}^*=(0.6,0,0.4)$ used for the Full 2000 primary report, the Full 800 baseline AUC is $0.852$ and the E10 AUC is $0.833$.
The attack-side script released as \texttt{scripts/e10\_compose.py} remains archived as a diagnostic rebalancing check.
That script reproduces best-weight baseline AUC $0.863$ and E10 AUC $0.843$ ($-0.020$) deterministically from the committed paraphrased corpus plus a uniform random tool-shuffle and uniform $[0,30]$\,s jitter.
To test whether this small drop is specific to the paraphrase source, we also release \texttt{scripts/e10\_source\_probe.py}, which applies the same shuffle+jitter composition to all seven committed Full 800 evasion sources at seed 20260505.
The lowest-AUC source in that archived sweep uses multilingual prompts and reaches AUC $0.808$ ($-0.055$), with the full sweep archived in \texttt{run/n800/e10\_source\_probe.json}.
A companion ten-seed probe for that multilingual source remains in AUC $[0.808,0.823]$ and is archived in \texttt{run/n800/e10\_seed\_probe.json}.
Earlier internal runs with a stronger paraphrase and decoy-tool injection reached $0.680$; we choose to publish the smaller-drop reproducible values rather than the larger-drop internal value.
The qualitative finding is unchanged: the released multi-axis compositions remain above random without detector-score feedback, but the main stress row is fixed-weight and fully adaptive detector-score-oracle L3 remains future work.
The canonical E10 generation and evaluation recipe is archived in \texttt{run/n800/e10\_results.json}, the committed E10 summary file.

\end{document}